\documentclass[journal]{IEEEtran}
\usepackage{graphicx,amsmath,array,subfigure,url,booktabs,xcolor,multirow,color,balance,amsfonts,algorithmic,algorithm,textcomp,stfloats,verbatim,newclude,accents,amssymb,amsthm,cite,placeins,bm,threeparttable,multirow}
\usepackage{tabularx}
\usepackage{array}
\usepackage{amssymb}  
\usepackage{pifont}   

\renewcommand{\arraystretch}{1.3} 
\renewcommand{\multirowsetup}{\centering} 
\newtheorem{remark}{Remark}
\newtheorem{theorem}{Theorem}

\begin{document}

\title{Polarforming for Wireless Communications: \\Modeling and Performance Analysis}

\author{
	Zijian Zhou, \IEEEmembership{Member,~IEEE},
	Jingze Ding, \IEEEmembership{Graduate Student Member,~IEEE},
  Chenbo Wang, \IEEEmembership{Member,~IEEE},\\
  Bingli Jiao, \IEEEmembership{Senior Member,~IEEE}, and
  Rui Zhang, \IEEEmembership{Fellow,~IEEE}
	\thanks{The work was supported in part by the National Natural Science Foundation of China under grants No. 62471424, the Basic Research Project No. HZQB-KCZYZ-2021067 of Hetao Shenzhen-HK S\&T Cooperation Zone, the National Natural Science Foundation of China under grants No. 92267202 and U25A20390, the Shenzhen Fundamental Research Program under grant No. JCYJ20250604141209012, The Guangdong Provincial Key Laboratory of Big Data Computing, the National Natural Science Foundation of China under grants No. 62331022, and the Guangdong Major Project of Basic and Applied Basic Research under grants No. 2023B0303000001. (\textit{Corresponding author: Rui Zhang.})}
	\thanks{Zijian Zhou is with the School of Science and Engineering, The Chinese University of Hong Kong, Shenzhen 518172, China (e-mail: zijianzhou@link.cuhk.edu.cn).}
	\thanks{Jingze Ding and Chenbo Wang are with the School of Electronics, Peking University, Beijing 100871, China (e-mail: djz@stu.pku.edu.cn; wcb15@pku.edu.cn).}
	\thanks{Bingli Jiao is with the School of Computing and Artificial Intelligence, Fuyao University of Science and Technology, Fuzhou 350109, China.  He is also with the School of Electronics, Peking University, Beijing 100871, China (e-mail: jiaobl@pku.edu.cn).}
	\thanks{Rui Zhang is with the Department of Electrical and Computer Engineering, National University of Singapore, Singapore 117583 (e-mail: elezhang@nus.edu.sg).}}
\maketitle

\begin{abstract}
This paper presents the novel concept of \textit{polarforming} for wireless communications, which enables the polarization of an antenna to shape into a desired polarization state for aligning with the polarization of an electromagnetic (EM) wave.  Through polarization matching, it can fully leverage polarization diversity and mitigate polarization loss to enhance the performance of wireless communication systems.  To implement polarforming, we propose a new paradigm of phase shifter (PS)-based polarization-reconfigurable antennas (PRAs) that can form linear, circular, and general elliptical polarizations by continuous phase shift control.  To further demonstrate the benefits of polarforming, we investigate a PS-based PRA aided wireless communication system equipped with tunable polarization of antennas.  We characterize the multiple-input multiple-output (MIMO) channel capacity of the considered system as a function of the phase shifts of PS-based PRAs.  We also provide a detailed polarforming interpretation under the single-input single-output (SISO) scenario and theoretically show how polarforming differs from the conventional (analog) beamforming based on PSs.  Moreover, we develop an alternating optimization approach to maximize the channel capacity for systems with a single-antenna transmitter/receiver.  Based on the water-filling principle, we also derive an upper bound on the MIMO channel capacity with PS-based PRAs and then maximize this capacity bound by optimizing the phase shifts through alternating optimization.  Finally, comprehensive simulation results are presented, which not only validate the effectiveness of polarforming in combating channel depolarization but also exhibit substantial performance improvements over conventional systems.
\end{abstract}

\section{Introduction} \label{sec1}
\IEEEPARstart{P}{olarization} describes the characteristics of an electromagnetic (EM) wave, and theoretically, it holds the potential to triple the wireless channel capacity, by exploiting the existence of six distinct electric and magnetic polarization states at any given point \cite{ref_AMD01}.  Nonetheless, classical wireless systems place an overfull focus on optimizing time, frequency, and spatial resources, often neglecting polarization as another dimension \cite{ref_AOK24, ref_RXK19, ref_LLS14}.  Although these systems have been successfully implemented, they are now reaching their theoretical and practical performance boundaries.  To meet the unprecedented demands for higher data rates and more reliable connections, leveraging polarization diversity in wireless systems is becoming a promising and necessary evolution.

The polarization characteristics of wireless channels have been extensively investigated in the literature \cite {ref_MO09, ref_KS11, ref_Coldrey08, ref_HCS16, ref_NBE02, ref_ESB04, ref_ESC06, ref_ABH07, ref_VGM09}, which provide a solid foundation for understanding the impact of polarization on wireless communications.  Specifically, the authors in \cite{ref_MO09, ref_KS11} studied the mechanisms of channel depolarization in both line-of-sight (LoS) and non-line-of-sight (NLoS) scenarios from an electromagnetism perspective.  They discovered that reflections, diffractions, and scattering of EM waves, and even antenna misalignment can cause significant changes in polarization states, thereby leading to depolarization.  To quantify the degree of depolarization, cross-polarization discrimination (XPD) was introduced as the ratio of the average received power in the co-polar channel to that in the cross-polar channel \cite{ref_Coldrey08, ref_HCS16}.  The XPD provides a measurement of depolarization effects under different conditions.  Further studies \cite{ref_NBE02, ref_ESB04, ref_ESC06, ref_ABH07, ref_VGM09} measured the XPD in different environments and showed that channel depolarization can result in a notable capacity degradation in practical wireless communication systems.  Since the events that cause depolarization are random and highly dependent on the surrounding environment, the conventional systems with fixed-polarization antennas (FPAs) cannot effectively adapt to channel variations and mitigate channel depolarization, particularly in challenging scenarios like vehicular communications or unmanned aerial vehicle networks \cite{ref_YSL23, ref_MWX24}.

\renewcommand{\arraystretch}{1.3} 
\renewcommand{\multirowsetup}{\centering}
\begin{table*}[!t]
	\centering
	{
	\begin{threeparttable}
		\caption{Comparison of different polarized antenna architectures.}
		\label{tab1}
		\begin{tabular}{|c|c|c|c|c|c|}
			\hline
			{\bf Polarization dimension}& {\bf Antenna architecture}
			&{\bf RF chain} & {\bf Hardware complexity} & {\bf Signal processing complexity} & {\bf Performance gain}\\
			\hline
			\hline
			\multirow{2}{*}{Uni-polarized} & LPA & Single & Very low & Low & Very low \\
			\cline{2-6}
			& PAA \cite{ref_KM15, ref_OKY18, ref_OK20, ref_OKM21, ref_OHK24, ref_CH21, ref_CH24, ref_DKY19, ref_DKY21} & Single & Moderate & Moderate & Moderate \\
			\hline
			\multirow{4}{*}{Dual-polarized} & CPA & Single & Very low & Low & Low \\
			\cline{2-6}
			& SPRA \cite{ref_SJK04, ref_GSZ06, ref_KRR15, ref_ZCL15, ref_RZW20} & Single & Low & Moderate & Moderate \\
			\cline{2-6}
			& DPA \cite{ref_OB23, ref_OCG08, ref_Ertug08, ref_IKB24, ref_WPF19, ref_WPF20, ref_HXK20, ref_SCL15, ref_ZCH17} & Double & High & High & High \\
			\cline{2-6}
			& PS-based PRA & Single & Low & Moderate & High \\
			\hline
		\end{tabular}
	\end{threeparttable}
	}
\end{table*}

The dual-polarized antenna (DPA) is the most well-explored technology for leveraging polarization diversity and overcoming the inherent limitations of FPAs.  Tri-polarized antennas can also be utilized, but they are primarily suited for short-range communications where substantial scattering occurs around the transceiver \cite{ref_OB23}.  Therefore, tri-polarized antennas are not considered in this paper.  In DPA systems, each antenna has two ports with orthogonal polarization orientations and requires two dedicated radio frequency (RF) chains. Over the past decades, many studies, such as \cite{ref_OB23, ref_OCG08, ref_Ertug08, ref_IKB24}, have assessed and validated the performance of DPAs in terms of multiple-input multiple-output (MIMO) channel capacity.  Towards digital modulation, the studies \cite{ref_WPF19, ref_WPF20} proposed polarization shift keying modulation using DPAs, and demonstrated that this modulation achieves a lower error rate compared to unit symbol power constellations.  Although the authors in \cite{ref_HXK20} proposed a different polarization modulation encoder in the analog domain with the aid of DPAs, the implementation cost associated with switching between different predefined polarization states is still practically high.  Furthermore, the use of double RF chains in DPA systems considerably raises its cost, making it impractical for lightweight, low-complexity wireless devices in many applications such as the Internet of Things \cite{ref_XHL14}.  As a result, DPAs have been primarily deployed at base stations in cellular networks, but the cost remains prohibitive and may be practically unacceptable for future wireless networks such as millimeter-wave systems.  In \cite{ref_SCL15, ref_ZCH17}, hybrid beamforming was investigated with dual-polarized arrays with limited polarization adjustment.

To reduce the cost of DPAs, polarization-reconfigurable antennas (PRAs) have been developed, commonly known as switchable PRAs (SPRAs) due to their capability to switch between predefined polarization states.  The implementation of SPRAs is well-established, relatively low-cost, and benefits from advancements in antenna architecture.  { In the realm of antenna and propagation, various types of SPRAs were invented to switch among linear polarization, left-handed circular polarization, and right-handed circular polarization by using positive intrinsic negative (PIN) diodes \cite{ref_SJK04, ref_GSZ06}, micro-electromechanical systems (MEMS) switches \cite{ref_KRR15}, metasurfaces \cite{ref_ZCL15}, dielectric liquid \cite{ref_RZW20}, etc.  Nevertheless, SPRAs usually support a limited number of predefined polarization states, e.g., two in \cite{ref_KRR15, ref_RZW20} and four in \cite{ref_SJK04, ref_GSZ06, ref_ZCL15}.  This limitation constrains the ability of the SPRA system to fully leverage polarization diversity and combat channel depolarization.}  To achieve higher diversity orders in wireless transmission, it is required to incorporate more antenna elements for switchable polarization states, which in turn will cause increased cost and system complexity.

An alternative approach is employing polarization-agile antennas (PAAs) to overcome the discrete adjustment limitations of SPRAs while maintaining a single RF chain per antenna.  The PAA system enables continuous adjustment of polarization orientation through the use of linearly polarized antennas (LPAs) within a given antenna panel.  There are several preliminary studies in the literature that explored the application of PAAs for wireless communications \cite{ref_KM15, ref_OKY18, ref_OK20, ref_OKM21, ref_OHK24, ref_CH21, ref_CH24, ref_DKY19, ref_DKY21}.  Specifically, the authors in \cite{ref_KM15} first proposed a polarization pre-post coding scheme assisted by PAAs and developed an iterative solution to find optimal transmit and receive polarization vectors for the purpose of maximizing channel capacity.  In \cite{ref_OKY18}, the authors evaluated the improvement of the PAA-aided MIMO system over the FPA system in capacity under different types of wireless channels.  In \cite{ref_OK20}, the authors proposed an energy-efficient PAA system and developed a multi-polarization superposition beamforming optimization framework under the fifth-generation (5G) antenna panel structure.  The results showed that the system is XPD-aware for transmit power allocation.  The integration of antenna selection and PAAs was studied in \cite{ref_OKM21} to enhance channel capacity by further exploiting spatial and polarization diversity.  The authors in \cite{ref_OHK24} focused on minimizing pilot overhead by utilizing deep neural networks at the transmitter and receiver.  Instead of relying on explicit channel estimation, the proposed approach optimized polarization and beamforming vectors directly based on the received pilot signals.  The authors in \cite{ref_CH21, ref_CH24} considered the antenna pattern of the PAA system in wideband communications, where the reconfigurable array serves as polarization precoding/combining to optimize polarization vectors over unpolarized channels.  Moreover, the PAA system allows for dynamic polarization adjustment, but its implementation poses significant challenges.  The authors in \cite{ref_DKY19, ref_DKY21} proposed the use of discrete polarization angles controlled by RF switches to equivalently adjust antenna orientations.  However, this approach leads to performance degradation and encounters similar limitations as SPRAs.  In addition, PAAs can mitigate channel depolarization to some extent; however, the performance is limited by the use of LPAs, instead of circularly polarized antennas (CPAs) \cite{ref_DGM17}.

It is worth clarifying the distinction between polarforming and the polarization precoding/combining schemes.  The latter perform polarization adaptation in the baseband digital domain by weighting the outputs of dual-polarized ports, which requires a dedicated RF chain for each port.  In contrast, polarforming directly shapes the radiated polarization in the analog domain at the antenna front-end, where a single PS continuously controls the phase difference between the two orthogonal elements of a PRA driven by only one RF chain.  Polarforming is also fundamentally different from the SPRA, which can only switch among a limited number of predefined polarization states and thus cannot fully leverage polarization diversity, as discussed above.  This work thus introduces the PS-based PRA as a new hardware paradigm that realizes continuous polarization control with a single RF chain, establishes a unified channel model that expresses the MIMO capacity explicitly as a function of the PS phases, and characterizes the capacity limit together with low-complexity solutions and a capacity upper bound.  The architectural distinctions and performance advantages of the proposed PS-based PRA over existing alternatives are further summarized in Table \ref{tab1}.

In light of the above, this paper proposes a new PRA design based on phase shifter (PS) for wireless communication systems.  Different from the aforementioned PRAs, the PS-based PRA enables continuous polarization through polarization phase control at a very low implementation cost.  Furthermore, the PS-based PRA outperforms the conventional FPA and PRA in adapting to channel variations and combating channel depolarization under typical polarized channel conditions.  { Table \ref{tab1} summarizes the architectural distinctions and performance advantages of the proposed PS-based PRA against conventional alternatives.}  

More specifically, the main contributions of this paper are summarized as follows.
\begin{itemize}
	\item We propose a new design of PS-based PRAs, where each PRA can independently control the phase difference of vertical and horizontal polarizations to form linear, circular, and general elliptical polarizations.  Based on this framework, we investigate a single-user MIMO system with PS-based PRAs and formulate the MIMO channel capacity as a function of the phase shifts of the transmit and receive PRAs.
	
	\item We introduce the novel concept of \textit{polarforming} for wireless communications.  Polarforming is a form of waveforming in the polarization domain, where the wireless system leverages PRAs to shape the polarization of EM waves into a desired polarization state.  Polarforming also provides a physical interpretation of how PRAs can fully exploit polarization diversity.  It allows PRAs to serve as a feasible alternative to DPAs by reducing the required RF chains by half while preserving the full use of polarization diversity in wireless communication systems.
	
	\item We derive an upper bound of the MIMO channel capacity with respect to the phase shifts of the transmit and receive PS-based PRAs.  To maximize the channel capacity, we develop an alternating optimization approach to maximize this upper bound by iteratively optimizing the phase shifts while accounting for practical system constraints.  The proposed algorithm exhibits excellent convergence behavior and low computational complexity, which make it well-suited for practical applications.
	
	\item We carry out comprehensive simulations to evaluate the performance gains achieved by polarforming under different channel conditions.  The results demonstrate that the proposed system can effectively mitigate channel depolarization and offer substantial improvements over conventional systems.  Furthermore, the proposed system even outperforms the DPA system with double RF chains in the low signal-to-noise ratio (SNR) regime because of the additional noise introduced by more RF chains at the receiver in the DPA system.
\end{itemize}

The rest of this paper is organized as follows.  In Section \ref{sec2}, we introduce the principles of EM wave propagation and propose the system model for the PRA-aided wireless communication system.  In Section \ref{sec3}, we provide a physical interpretation of polarforming and develop an alternating optimization approach to maximize the upper bound of the MIMO channel capacity.  Section \ref{sec4} presents numerical results to validate the performance gains of the proposed system, and this paper is concluded in Section \ref{sec5}.

\begin{figure*}[!t]
	\centering
	\includegraphics[width=\linewidth]{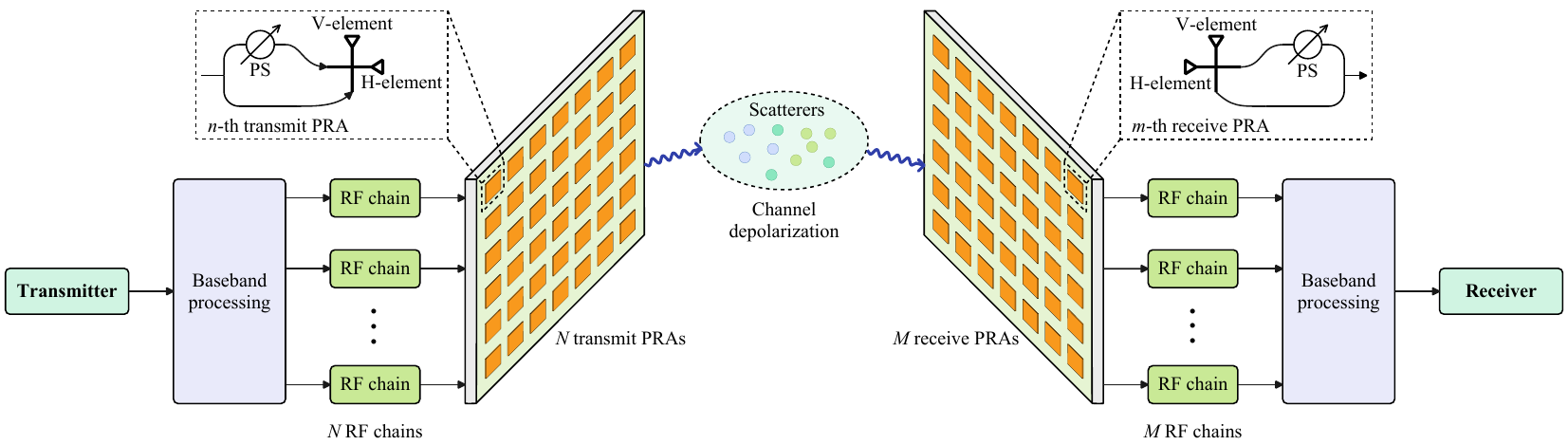}
	\caption{A schematic for the wireless communication system with PS-based PRAs.}
	\label{fig_system_model}
\end{figure*}

\textit{Notations}: $a$, $\mathbf{a}$, $\mathbf{A}$, and $\mathcal{A}$ denote a scalar, a vector, a matrix, and a set, respectively.  $(\cdot)^*$, $(\cdot)^T$, and $(\cdot)^H$ denote the conjugate, transpose, and conjugate transpose, respectively.  $\mathbb{E}\{\cdot\}$ denotes the mathematical expectation of a random variable.  $\left\|\cdot\right\|_2$ denotes the Euclidean norm of a complex-valued vector, and $\odot$ denotes the Hadamard product.  $\mathbb{R}^{p\times q}$ and $\mathbb{C}^{p\times q}$ represent the spaces of $p\times q$ real-valued and complex-valued matrices, respectively.  $\angle{x}$ and $|x|$ denote the phase and absolute value of a complex number $x$, respectively.  $\max(a,b)$ and $\min(a,b)$ select the maximum and minimum values between $a$ and $b$, respectively.    For a matrix ${\bf A}$, $[{\bf A}]_{pq}$ denotes its entry in the $p$-th row and $q$-th column, $\det({\bf A})$ denotes its determinant, ${\rm Tr}({\bf A})$ denotes its trace, and $\mathbf{A}^{1/2}$ denotes its square root.  ${\bf A} \succeq {\bf 0}$ indicates that ${\bf A}$ is a positive semi-definite matrix.  ${\bf B} = {\rm diag}\{x_1,x_2,\cdots,x_L\}$ denotes a diagonal matrix with diagonal entries $x_1,x_2,\cdots,x_L$, and ${\bf C} = {\rm blkdiag}\{{\bf x}_1,{\bf x}_2,\cdots,{\bf x}_L\}$ denotes a block diagonal matrix composed of the blocks ${\bf x}_1,{\bf x}_2,\cdots,{\bf x}_L$ along the diagonal.  $\mathbf{I}_K$ denotes an identity matrix of order $K$.  $\in$,  $\triangleq$, and $\sim$ stand for ``belong(s) to'', ``defined as'', and ``distributed as'', respectively.  $\mathcal{CN}({\bf 0},{\bf \Lambda})$ represents the circularly symmetric complex Gaussian (CSCG) distribution with mean zero and covariance matrix ${\bf \Lambda}$.

\section{System Model and Problem Formulation} \label{sec2}
\subsection{System Model}
As shown in Fig.\;\ref{fig_system_model}, we consider a wireless communication system assisted by $N$ transmit and $M$ receive PS-based PRAs.  Each PRA is equipped with only a single RF chain and consists of two orthogonal antenna elements\footnote{While the PS-based PRA in this paper employs two antenna elements, it is worth noting that our model can be extended to accommodate three orthogonal antenna elements by incorporating a PS with the additional antenna element.}, i.e., V-element for vertical polarization and H-element for horizontal polarization.  A PS is integrated into each antenna front-end to control the phase difference between the two antenna elements for continuous polarization control, with its tunable range assumed to span from $0$ to $2\pi$.  We denote $\theta_n$ as the phase shift of the $n$-th transmit PRA and $\phi_m$ as the phase shift of the $m$-th receive PRA, where $n=1,2,\cdots,N$ and $m=1,2,\cdots,M$.   These phase shifts are aggregated into phase shift vectors (PSVs), defined as
\begin{align}
	{\bm \vartheta} &= \left[ \theta_1, \theta_2, \cdots, \theta_N \right]^T \in \mathcal{A}, \\
	{\bm \varphi} &= \left[ \phi_1, \phi_2, \cdots, \phi_M \right]^T \in \mathcal{B}, 
\end{align}
where the sets $\mathcal{A} \triangleq [0,2\pi]^N$ and $\mathcal{B} \triangleq [0,2\pi]^M$.

For the proposed MIMO system, we consider narrow-band quasi-static channels, and the channel matrix can be expressed as a function of the phase shifts, i.e., ${\bf H}({\bm \vartheta}, {\bm \varphi})\in \mathbb{C}^{M\times N}$.  Let ${\bf s} \in \mathbb{C}^N$ denote the transmit signal vector with mean zero and covariance matrix ${\bf Q} \triangleq \mathbb{E} \{{\bf s} {\bf s}^H\} \in \mathbb{C}^{N\times N}$, where ${\bf Q} \succeq {\bf 0}$.  We consider an average sum power constraint at the transmitter given by $\mathbb{E}\{ \| {\bf s} \|^2_2\} \le P_{\rm t}$, which is equivalent to ${\rm Tr}({\bf Q}) \le P_{\rm t}$.  Therefore, the received signal vector of the PRA system can be written as
\begin{equation}
	{\bf y}({\bm \vartheta},{\bm \varphi}) = {\bf H}({\bm \vartheta},{\bm \varphi}) {\bf s} + {\bf z},
\end{equation}
where ${\bf z} \sim {\cal CN}({\bf 0}, \sigma^2 {\bf I}_M)$ denotes the additive white Gaussian noise (AWGN) vector at the receiver, with an average noise power of $\sigma^2$.  We assume that the noise from antenna elements and PSs is negligible compared to that from the RF chains, as the former is typically much lower than the latter introduced by active RF chain components.

Next, we define polarforming vectors (PFVs) in terms of phase shifts to characterize the polarization of PS-based PRAs as follows
\begin{equation} \label{PFV}
	{\bf f}(\theta_n) \triangleq \frac{1}{\sqrt{2}} \begin{bmatrix} 1\\e^{j\theta_n} \end{bmatrix} \;\text{and}~{\bf g}(\phi_m) \triangleq \begin{bmatrix} 1\\e^{j\phi_m} \end{bmatrix}.
\end{equation}
The factor $\frac{1}{\sqrt{2}}$ ensures normalization to comply with the transmit power constraint.  The transmit and receive PFVs in \eqref{PFV} capture the relationship between phase shifts and polarization configurations of the transmit and receive PRAs.  To extend this to the MIMO system, we define polarforming matrices (PFMs) by aggregating individual PFVs into block diagonal matrices as follows
\begin{align}
	{\bf F}({\bm \vartheta}) &\triangleq {\rm blkdiag}\left\{ {\bf f}(\theta_1), {\bf f}(\theta_2), \cdots, {\bf f}(\theta_N) \right\} \in \mathbb{C}^{2N\times N}, \\
	{\bf G}({\bm \varphi}) &\triangleq {\rm blkdiag}\left\{ {\bf g}(\phi_1), {\bf g}(\phi_2), \cdots, {\bf g}(\phi_M) \right\} \in \mathbb{C}^{2M\times M}.
\end{align}

Using these definitions, the overall channel matrix can be expressed as ${\bf H}({\bm \vartheta}, {\bm \varphi}) \triangleq {\bf G}({\bm \varphi})^H {\bf P} {\bf F}({\bm \vartheta})$, where ${\bf P}$ is a polarized channel matrix, given by \cite{ref_HCS16}
\begin{equation} \label{PCM}
	\mathbf{P} = \begin{bmatrix}
		{\bf P}_{11} & \cdots & {\bf P}_{1N} \\
		\vdots & \ddots & \vdots \\
		{\bf P}_{M1} & \cdots & {\bf P}_{MN}
	\end{bmatrix} \in \mathbb{C}^{2M \times 2N}.
\end{equation}
Each submatrix ${\bf P}_{mn}$ in \eqref{PCM} represents the two-by-two polarized channel matrix between the $n$-th transmit PRA and the $m$-th receive PRA.

\subsection{Problem Formulation}
To explore the capacity limit of the proposed MIMO system, we assume that perfect channel state information (CSI) of the polarized channel matrix ${\bf P}$ in \eqref{PCM} is available at both the transmitter and receiver\footnote{{ Acquiring the CSI of the polarized channel matrix is challenging, as discussed in Section \ref{sec3-4}, while we make this assumption in order to derive the capacity limit of the system under investigation.  Besides, the impact of polarized channel estimation error is studied later in Section \ref{sec4}.}}.  Under this assumption, the MIMO channel capacity in bits-per-second-per-Hertz (bps/Hz) is given by
\begin{align} \label{capacity_mimo}
	C(&{\bm \vartheta}, {\bm \varphi}) \nonumber\\
	&= \max_{{\substack {{\bf Q}: {\bf Q} \succeq \mathbf {0}, \\ {\rm Tr}({\bf Q}) \leq P_{\rm t}}} } \log_2\det \left( {\bf I}_M + \frac{1}{\sigma^2} {\bf H}({\bm \vartheta}, {\bm \varphi}) {\bf Q} {\bf H}({\bm \vartheta}, {\bm \varphi})^H \right).
\end{align}
Unlike conventional FPA systems, the channel capacity of the PS-based PRA aided MIMO system in \eqref{capacity_mimo} depends on the phase shifts of the transmit and receive PRAs.  These phase shifts determine the polarization configurations and also influence the optimal transmit covariance matrix ${\bf Q}$ for the channel matrix ${\bf H}({\bm \vartheta}, {\bm \varphi})$.

In this paper, we aim to maximize the channel capacity of the PRA-aided MIMO system by jointly optimizing the transmit and receive PSVs, ${\bm \vartheta}$ and ${\bm \varphi}$, and the transmit covariance matrix ${\bf Q}$ subject to the phase shift and transmit power constraints.  Accordingly, the optimization problem can be formulated as
\begin{subequations}
\label{opt_problem}
\begin{align}
		\max_{ {\bm \vartheta}, {\bm \varphi}, {\bf Q}} ~ & ~ \log_2\det \left( {\bf I}_M + \frac{1}{\sigma^2} {\bf H}({\bm \vartheta},{\bm \varphi}) {\bf Q} {\bf H}({\bm \vartheta},{\bm \varphi})^H \right)  \\
		{\rm s.t.} ~ & ~ {\bm \vartheta} \in \mathcal{A}, \label{cons_theta} \\
		& ~ {\bm \varphi} \in \mathcal{B}, \label{cons_phi} \\
		& ~ {\bf Q} \succeq {\bf 0}, \\
		& ~ {\rm Tr}({\bf Q}) \leq P_{\rm t}. 
\end{align}
\end{subequations}
Problem \eqref{opt_problem} is difficult to solve because the objective function is highly non-concave with respect to ${\bm \vartheta}$ and ${\bm \varphi}$.  Additionally, the coupling between the optimization variables ${\bm \vartheta}$, ${\bm \varphi}$, and ${\bf Q}$ adds to the complexity of the problem.  To address these challenges, we will develop an alternating optimization approach in the subsequent section, which iteratively optimizes one set of variables while keeping the others fixed, thereby simplifying the problem into more tractable subproblems \cite{ref_ZZ20, ref_MZZ23}.

\section{Performance Analysis and Proposed Solution} \label{sec3}
In this section, we provide a detailed illustration of polarforming as well as an explanation of its differences from the conventional (analog) beamforming.  Based on an alternating optimization approach, we then propose the solutions for capacity maximization in multiple-input single-output (MISO), single-input multiple-output (SIMO), and MIMO systems, followed by a discussion of practical considerations.

\subsection{Polarforming Interpretation} \label{sec3-1}
To better explore the properties of polarforming, we start by considering a single-input single-output (SISO) system equipped with an FPA at the transmitter and a PS-based PRA at the receiver to perform receive polarforming.  The channel response of this SISO system is given by $h(\phi) = {\bf g}(\phi)^H {\bf b}$, where ${\bf b} \triangleq [b_1, b_2]^T \in \mathbb{C}^2$ is referred to as an effective polarization state vector (EPSV), and $\phi$ is the phase shift of the receive PRA.  We note that the EPSV directly determines the polarization of the incoming EM waves.  Therefore, for a given EPSV, the channel power gain can be derived as
\begin{align} \label{h_phi2}
	|h(\phi)|^2 &= \left| b_1 + b_2 e^{-j\phi} \right|^2 \nonumber\\
	& = |b_1|^2 + |b_2|^2 + 2|b_1||b_2| \cos(\phi + \angle{b_1} - \angle{b_2}) \nonumber \\
	& \mathop \le \limits^{(a)} \left(|b_1| + |b_2|\right)^2.
\end{align}
The equality marked by $(a)$ holds when the cosine term reaches its maximum value of one.  Thus, the optimal phase shift $\phi$ to maximize the channel power gain is given by
\begin{equation}
	\phi^\star = \angle{b_2} - \angle{b_1}.
\end{equation}
While the above configuration is based on receive polarforming, transmit polarforming can be operated in a similar manner.  It can be observed that the optimal transmit/receive polarforming design depends only on the phase difference of the V-element and H-element of the PRA.

\begin{figure}[!t]
	\centering
	\includegraphics[width=\linewidth]{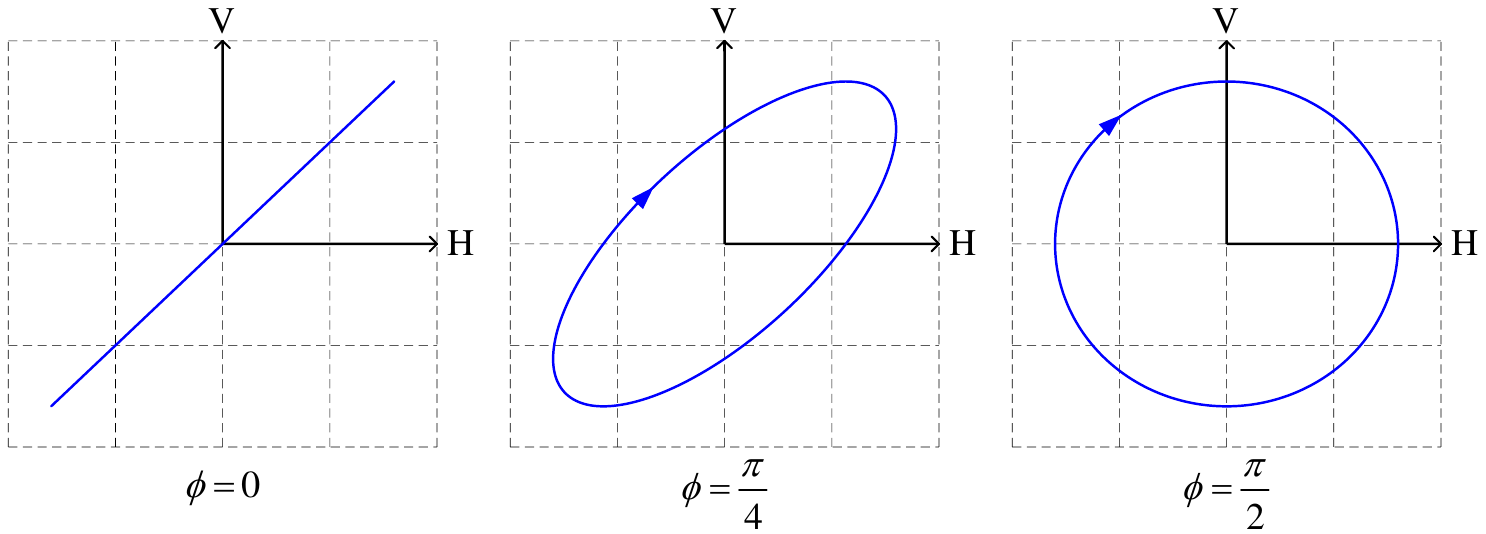}
	\caption{An example of the polarization of a PS-based PRA for different phase shifts.}
	\label{fig_antenna_polarization}
\end{figure}

\begin{figure}[!t]
	\centering
	\includegraphics[width=\linewidth]{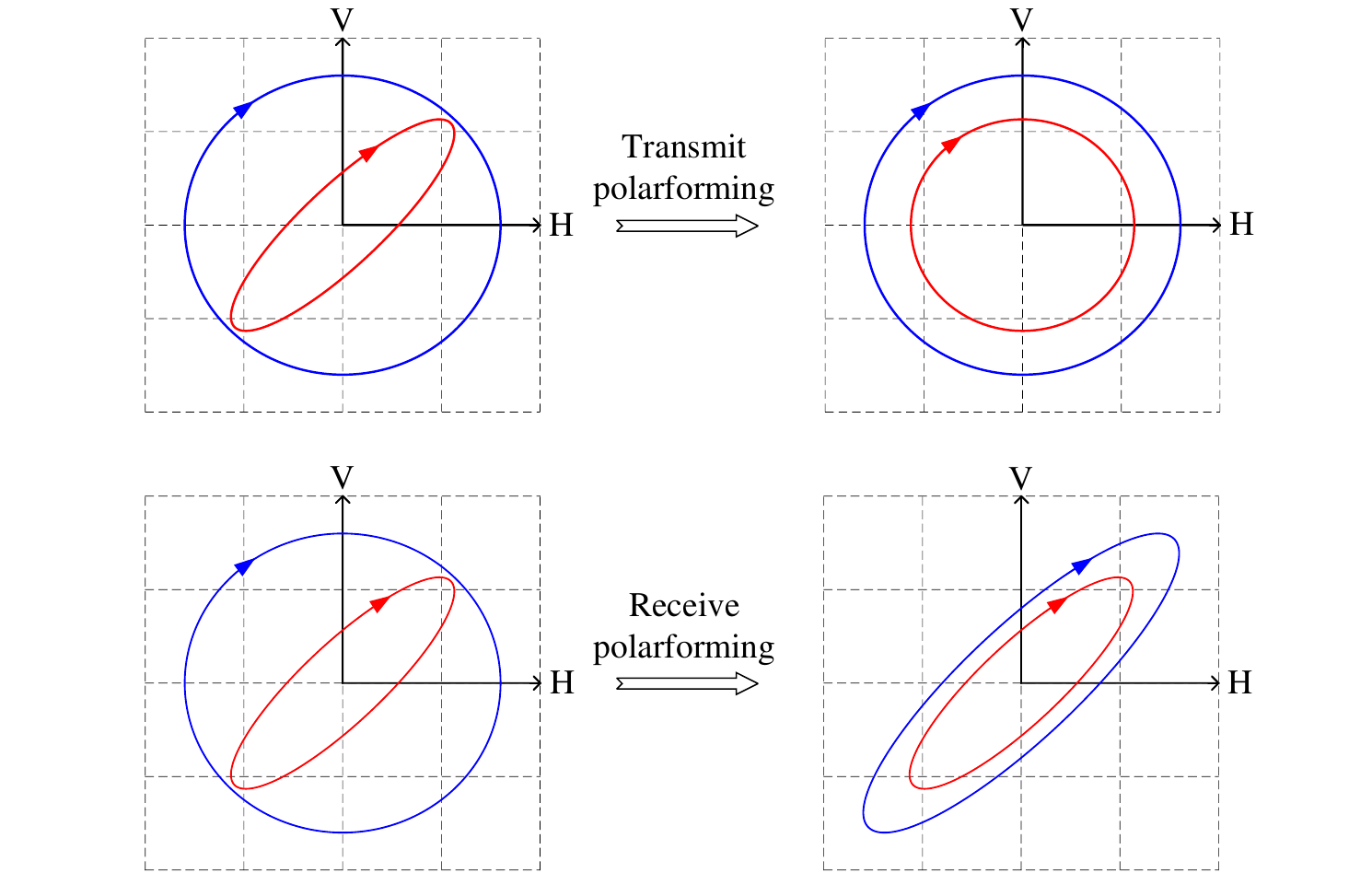}
	\caption{Illustration of transmit/receive polarforming via planes of polarization (PoPs). The blue line shows the polarization of the receive antenna, while the polarization of the incoming EM wave is highlighted in red.}
	\label{fig_polarforming}
\end{figure}

{ We illustrate the polarization of the proposed PRA using the receive PFV (determined by the phase shift $\phi$), as shown in Fig.\;\ref{fig_antenna_polarization}.  We observe that modifying the phase shift enables the formation of linear, circular, and general elliptical polarizations.  This capability accordingly allows for the modification of the polarization of transmitted/received EM waves.  To further illustrate this, an example is provided in Fig.\;\ref{fig_polarforming} to demonstrate transmit/receive polarforming from the perspective of polarization states.  In this figure, the original EPSV (without polarforming) is fixed with $|b_1| = |b_2| = \frac{\sqrt{2}}{2}$ and $\angle{b_1} - \angle{b_2} = -\frac{\pi}{6}$.  The left side of the figure depicts the effective polarization of the incoming EM wave (left-handed elliptical polarization) and the polarization of the receive antenna (left-handed circular polarization).  The right side displays these polarization states achieved by polarforming.  It can be seen that transmit polarforming achieves polarization matching by modifying the polarization of the wave to align it with the polarization of the receive antenna.  Similarly, receive polarforming adjusts the polarization of the receive antenna to align it with the polarization of the incoming EM wave.  This polarization matching ensures the maximum channel power gain in \eqref{h_phi2} for the PS-based PRA aided wireless communication system.}

{ The effectiveness of polarforming can be interpreted through the example above.  However, this is fundamentally distinct from the conventional PS-based beamforming, despite the similarities in their schematic representations.  Therefore, it is inappropriate to use the term \textit{beamforming} to describe this form of waveforming in the polarization domain, which is why we introduce the term \textit{polarforming}.  The following remarks are provided to distinguish polarforming from beamforming in more details.}

\begin{remark} \label{remark1}
Polarforming is fundamentally different from beamforming.  In beamforming, antennas are typically spaced at half a wavelength or more, thereby allowing phase shifts to steer the beam in a desired direction.  Nevertheless, for a PS-based PRA, the antenna elements used for polarforming are orthogonally polarized and spatially overlapping without physical separation.  Consequently, adjusting the phase shifts of these elements does not affect the beam's direction.
\end{remark}

\begin{figure}[!t]
	\centering
	\subfigure[Radiation pattern of a spatially separated two-element array ($d = \lambda/2$) with different inter-element phase shifts $\alpha$]{\label{fig_pattern_a}
		\includegraphics[width=0.6\columnwidth]{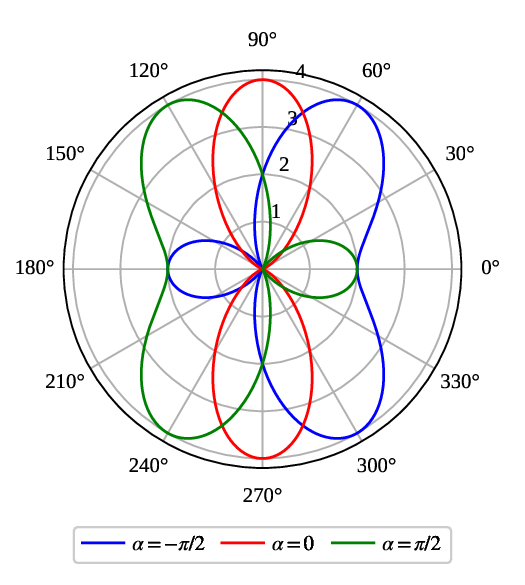}}
	\\
	\subfigure[Radiation pattern of the co-located PS-based PRA with different phase shifts $\phi$]{\label{fig_pattern_b}
		\includegraphics[width=0.6\columnwidth]{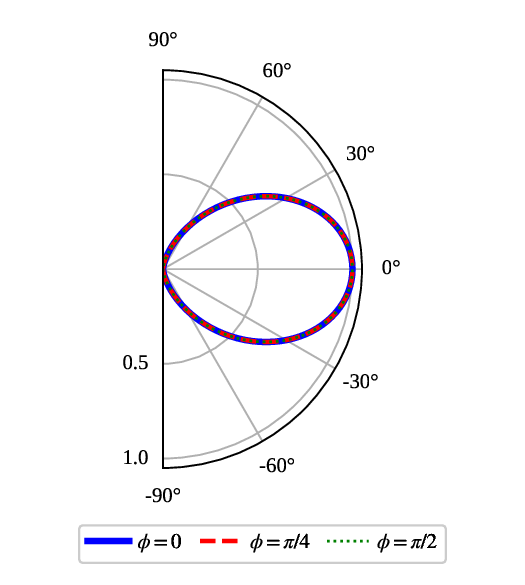}}
	\caption{Comparison between beamforming with a spatially separated array and polarforming with a co-located PS-based PRA.  The inter-element phase shifts steer the radiation pattern of the array, whereas the polarforming phase shift only reshapes the polarization state of the PRA without changing its radiation pattern.}
	\label{fig_pattern}
\end{figure}

To rigorously justify Remark \ref{remark1}, we examine the far-field radiation of the PS-based PRA.  Specifically, the PRA is modeled as two co-located antenna elements that share an identical element power pattern $G_{\rm e}(\theta_{\rm s}, \varphi_{\rm s})$ but have mutually orthogonal polarization unit vectors $\hat{\bf e}_{\rm V}(\theta_{\rm s},\varphi_{\rm s})$ and $\hat{\bf e}_{\rm H}(\theta_{\rm s},\varphi_{\rm s})$, where $\theta_{\rm s}$ and $\varphi_{\rm s}$ denote the elevation and azimuth angles of the spatial direction, respectively.  When the PRA is driven by the PFV ${\bf g}(\phi) = [1, e^{j\phi}]^T$, the radiated far field can be expressed as ${\bf E}(\theta_{\rm s},\varphi_{\rm s}) = \sqrt{G_{\rm e}(\theta_{\rm s},\varphi_{\rm s})}\,(\hat{\bf e}_{\rm V} + e^{j\phi}\hat{\bf e}_{\rm H})$, and the corresponding spatial power pattern is given by
\begin{align} \label{power_pattern}
	U(\theta_{\rm s},\varphi_{\rm s}) &= \left\| {\bf E}(\theta_{\rm s},\varphi_{\rm s}) \right\|_2^2 \nonumber\\
	&= G_{\rm e}(\theta_{\rm s},\varphi_{\rm s}) \left( \left\|\hat{\bf e}_{\rm V}\right\|_2^2 + \left\|\hat{\bf e}_{\rm H}\right\|_2^2 + 2\,{\rm Re}\!\left\{ e^{j\phi}\, \hat{\bf e}_{\rm V}^H \hat{\bf e}_{\rm H} \right\} \right) \nonumber\\
	&= 2\, G_{\rm e}(\theta_{\rm s},\varphi_{\rm s}),
\end{align}
where the last equality follows from the pointwise orthogonality $\hat{\bf e}_{\rm V}^H \hat{\bf e}_{\rm H} = 0$ and the unit norms $\|\hat{\bf e}_{\rm V}\|_2 = \|\hat{\bf e}_{\rm H}\|_2 = 1$.  Since $U(\theta_{\rm s},\varphi_{\rm s}) = 2 G_{\rm e}(\theta_{\rm s},\varphi_{\rm s})$ is independent of $\phi$, the spatial power pattern, and hence the beam direction, is invariant to the polarforming phase shift $\phi$.  By antenna reciprocity, this invariance holds for both transmission and reception.  In practice, the two elements of a PRA may not be ideally orthogonal due to the finite cross-polar isolation (XPI), in which case the relative variation of $U(\theta_{\rm s},\varphi_{\rm s})$ is bounded by the small cross-polar coupling $|\hat{\bf e}_{\rm V}^H \hat{\bf e}_{\rm H}|$, and the impact of imperfect antenna XPI on the achievable rate will be evaluated in Section \ref{sec4}.  The phase shift $\phi$ instead rotates the Jones vector $\hat{\bf e}_{\rm V} + e^{j\phi}\hat{\bf e}_{\rm H}$, i.e., it only reshapes the polarization state of the radiated EM wave without altering the angular power distribution.  In contrast, for beamforming with a spatially separated array, the array factor ${\bf w}^H {\bf a}(\theta_{\rm s})$ depends on the spatial direction through the steering vector ${\bf a}(\theta_{\rm s}) = [1, e^{j2\pi d\cos\theta_{\rm s}/\lambda}, \cdots]^T$, where ${\bf w}$ denotes the beamforming vector, $\lambda$ denotes the wavelength, and $d$ is the element spacing.  Adjusting the phase shifts in ${\bf w}$ thus moves the peak of the radiation pattern, i.e., steers the beam toward a desired direction.  For the PS-based PRA, this inter-element spatial phase vanishes since the two elements are co-located, i.e., $d = 0$.  Accordingly, beamforming exploits the spatial degree of freedom introduced by nonzero element spacing, whereas polarforming exploits the polarization degree of freedom introduced by the orthogonally polarized co-located elements.  Hence, they can be separately applied or jointly designed as elaborated in Remark \ref{remark2} below.  It is also worth distinguishing polarforming from near-field beamfocusing \cite{ref_ZSE22, ref_LMD23}, which concentrates energy at a specific spatial location, in both range and angle, by exploiting the spatial degrees of freedom offered by spherical wavefronts in the near field.  Such beamfocusing remains a spatial-domain operation and is thus also complementary to polarforming.  Fig.\;\ref{fig_pattern} numerically validates the above distinction.  As shown in Fig.\;\ref{fig_pattern_b}, varying $\phi$ leaves the radiation pattern of the PRA unchanged, while it continuously reshapes the polarization state, as illustrated earlier in Fig.\;\ref{fig_antenna_polarization}.  In contrast, varying the inter-element phase shift of a spatially separated array steers the beam toward different directions, as shown in Fig.\;\ref{fig_pattern_a}.

\begin{remark} \label{remark2}
Polarforming can be jointly designed with beamforming to enhance system performance.  In multi-PRA systems, where multiple PRAs are physically spaced apart, beamforming can be realized by carefully designing the phase shifts between the PRAs.  This spatial separation enables beam steering in the desired direction while polarforming by each PRA simultaneously aligns polarization states to combat channel depolarization.  Integrating both techniques can provide greater spatial and polarization diversity and hence lead to improved communication performance and robustness.
\end{remark}

\subsection{Capacity Maximization for MISO and SIMO Systems} \label{sec3-2}
In this subsection, we investigate two special cases for the considered system with a single-antenna transmitter/receiver and provide an alternating optimization solution for capacity maximization.

To begin with, we consider the MISO system assisted by PS-based PRAs for $N > 1$ and $M = 1$.  In this case, the receive PFV can be expressed as ${\bf g}(\phi)$, where $\phi$ is the phase shift of the receive PRA.  Then, the polarized channel matrix in \eqref{PCM} reduces to
\begin{equation}
	{\bf P} = \left[ {\bf P}_{11}, {\bf P}_{12}, \cdots, {\bf P}_{1N} \right] \in \mathbb{C}^{2\times 2N}.
\end{equation}
The effective channel row vector is given by ${\bf h}({\bm \vartheta}, \phi)^H = {\bf g}(\phi)^H {\bf P} {\bf F}({\bm \vartheta}) \in \mathbb{C}^{1\times N}$.  The optimal transmit covariance matrix in this MISO system can be obtained according to maximum ratio transmission (MRT), i.e., ${\bf Q}^\star = \frac{{\bf h}({\bm \vartheta}, \phi) {\bf h}({\bm \vartheta}, \phi)^H}{\left\| {\bf h}({\bm \vartheta}, \phi) \right\|^2_2}  P_{\rm t}$.  Thus, the channel capacity of the MISO system is expressed as
\begin{align} \label{capacity_miso}
	C_{\rm MISO}({\bm \vartheta}, \phi) &= \log_2\left( 1 + \frac{1}{\sigma^2} {\bf h} ({\bm \vartheta}, \phi)^H {\bf Q}^\star {\bf h} ({\bm \vartheta}, \phi) \right) \nonumber \\
	&= \log_2\left( 1 + \frac{P_{\rm t}}{\sigma^2} \left\| {\bf h} ({\bm \vartheta}, \phi) \right\|^2_2\right).
\end{align}

From \eqref{capacity_miso}, it can be observed that maximizing the channel capacity is equivalent to maximizing the channel power gain $\left\| {\bf h} ({\bm \vartheta}, \phi) \right\|^2_2$ subject to the transmit PSV ${\bm \vartheta}$ and the receive phase shift $\phi$.  Accordingly, the optimization problem is formulated as
\begin{subequations}
	\label{problem_miso}
	\begin{align}
		\max_{ {\bm \vartheta}, \phi} ~ & ~ \left\| {\bf h} ({\bm \vartheta}, \phi) \right\|^2_2 \\
		{\rm s.t.} ~ & ~ \eqref{cons_theta}, \\
		& ~ \phi \in [0,2\pi].
	\end{align}
\end{subequations}
We note that the objective function is non-concave due to the exponential terms in the PFVs.  Nonetheless, we can apply an alternating optimization approach to obtain a suboptimal solution iteratively, as shown in the following.

For given ${\bm \vartheta}$, the objective function of problem \eqref{problem_miso} is given by
\begin{equation} \label{h_phi}
	\left\|{\bf h}(\phi)\right\|^2_2 = {\bf g}(\phi)^H \left( \sum_{n = 1}^N {\bf P}_{1n} {\bf f}(\theta_n) {\bf f}(\theta_n)^H {\bf P}_{1n}^H \right)  {\bf g}(\phi).
\end{equation}
The matrix $\sum_{n = 1}^N {\bf P}_{1n} {\bf f}(\theta_n) {\bf f}(\theta_n)^H {\bf P}_{1n}^H$ in \eqref{h_phi} is Hermitian because each term in the sum is Hermitian.  With this observation, we introduce the following theorem to address problem \eqref{problem_miso}.

\begin{theorem} \label{theorem1}
	Let ${\bf W}$ be a $2\times 2$ Hermitian matrix, and define ${\bf p}(\psi) = [1, e^{j\psi}]^T$.  The optimal angle $\psi$ that maximizes the quadratic function ${\bf p}(\psi)^H {\bf W} {\bf p}(\psi)$ subject to $\psi \in [0,2\pi]$ is given by
	\begin{equation} \label{opt_psi}
		\psi^\star = \angle{[{\bf W}]_{21}}.
	\end{equation}
\end{theorem}

\begin{IEEEproof}
	Consider a Hermitian matrix ${\bf W}$ expressed as
	\begin{equation}
		{\bf W} = \begin{bmatrix}
			a & c^* \\
			c & d
		\end{bmatrix},
	\end{equation}
	where $a,d\in\mathbb{R}$ and $c\in\mathbb{C}$. The objective function can be rewritten as
	\begin{align} \label{psi_W_psi}
		{\bf p}(\psi)^H {\bf W} {\bf p}(\psi) &= ce^{-j\psi} + c^* e^{j\psi} + a + d \nonumber\\
		&= 2 |c| \cos(\psi-\angle{c}) + a + d.
	\end{align}

	To maximize the objective function ${\bf p}(\psi)^H {\bf W} {\bf p}(\psi)$, the cosine term in \eqref{psi_W_psi} must attain its maximum value of one for $\psi \in [0,2\pi]$.  This occurs when $\psi - \angle{c} = 0$, which implies $\psi^\star = \angle{[{\bf W}]_{21}}$.  Thus, the optimal angle in \eqref{opt_psi} is achieved, and the proof of the theorem is complete.
\end{IEEEproof}

According to Theorem \ref{theorem1}, the optimal phase shift $\phi$ to maximize the objective function $\left\|{\bf h}(\phi)\right\|^2_2$ in \eqref{h_phi} is then given by
\begin{equation} \label{opt_phi_miso}
\phi^\star = \angle{\left[ \sum_{n = 1}^N {\bf P}_{1n} {\bf f}(\theta_n) {\bf f}(\theta_n)^H {\bf P}_{1n}^H \right]_{21}}.
\end{equation}

For given $\phi$, the objective function in problem \eqref{problem_miso} can be rewritten as
\begin{equation} \label{h_vartheta}
	\left\|{\bf h}({\bm \vartheta})\right\|^2_2 = \sum_{n = 1}^N {\bf f}(\theta_n)^H \left( {\bf P}_{1n}^H {\bf g}(\phi) {\bf g}(\phi)^H {\bf P}_{1n} \right) {\bf f}(\theta_n).
\end{equation}
To maximize the objective function in problem \eqref{problem_miso}, it is equivalent to optimizing each phase shift of the transmit PRAs.  Thus, the optimal phase shift $\theta_n$ to maximize the objective function $\left\|{\bf h}({\bm \vartheta})\right\|^2_2$ in \eqref{h_vartheta} is given by
\begin{equation} \label{opt_theta_miso}
\theta_n^\star = \angle{\left[{\bf P}_{1n}^H {\bf g}(\phi) {\bf g}(\phi)^H {\bf P}_{1n}\right]_{21}},\;n = 1,2,\cdots,N.
\end{equation}
It can be inferred that for a fixed receive PRA, the optimal solution for the transmit PRAs is achieved by aligning the phase shifts through phase precoding when full CSI is available at the transmitter.  Actually, the phase shifts given in \eqref{opt_theta_miso} provide the optimal solution for adjusting the transmit PRAs with an FPA at the receiver, called transmit polarforming, which will be evaluated in Section \ref{sec4}.

\begin{algorithm}[!t]
	\caption{Proposed Solution for Solving Problem \eqref{problem_miso}}
	\label{alg1}
	\footnotesize
	\renewcommand{\algorithmicrequire}{\textbf{Input:}}
	\renewcommand{\algorithmicensure}{\textbf{Output:}}
	\begin{algorithmic}[1]
		\REQUIRE $N$, $P_{\rm t}$, $\sigma$, ${\bf P}$, $\epsilon_1$.
		\ENSURE ${\bm \vartheta}$, $\phi$.
		\STATE Initialize ${\bm \vartheta}^{(0)} = {\bf 0}_{N}$ and $\phi^{(0)} = 0$.
		\FOR{$i=1\rightarrow\hat{I}_{\max}$}
		\STATE Update ${\bm \vartheta}^{(i)} \leftarrow {\bm \vartheta}^{(i-1)}$ and $\phi^{(i)} \leftarrow \phi^{(i-1)}$.
		\STATE Calculate $\sum_{n = 1}^N {\bf P}_{1n} {\bf f}\left(\theta_n^{(i)}\right) {\bf f}\left(\theta_n^{(i)}\right)^H {\bf P}_{1n}^H$ in \eqref{h_phi}.
		\STATE Update $\phi^{(i)}$ according to \eqref{opt_phi_miso}.
		\STATE Calculate $\left\{{\bf P}_{1n}^H {\bf g}(\phi) {\bf g}(\phi)^H {\bf P}_{1n}\right\}_{n=1}^N$ in \eqref{h_vartheta}.
		\STATE Update ${\bm \vartheta}^{(i)}$ according to \eqref{opt_theta_miso}.
		\IF{Increase of the channel capacity in \eqref{capacity_miso} is below $\epsilon_1$}
		\STATE Break.
		\ENDIF
		\ENDFOR
		\STATE Set ${\bm \vartheta} = {\bm \vartheta}^{(i)}$.
		\STATE Set $\phi = \phi^{(i)}$.
		\RETURN ${\bm \vartheta}$, $\phi$.
	\end{algorithmic}
\end{algorithm}

Based on the above, Algorithm \ref{alg1} provides an iterative method to solve the optimization problem \eqref{problem_miso} suboptimally.  It initializes the variables ${\bm \vartheta}^{(0)} = {\bf 0}_N$ and $\phi^{(0)} = 0$, and iterates until a maximum number of iterations $\hat{I}_{\max}$ is reached, where ${\bf 0}_N$ denotes a zero vector of length $N$.  In each iteration, the algorithm computes the summation of matrix products in \eqref{h_phi} and \eqref{h_vartheta}, and subsequently updates the phase shifts ${\bm \vartheta}^{(i)}$ and $\phi^{(i)}$.  The process continues until the capacity increase falls below $\epsilon_1$ or the maximum number of iterations is reached.  The final values of ${\bm \vartheta}^{(i)}$ and $\phi^{(i)}$ are returned as the solution.

{ The convergence and computational complexity of Algorithm \ref{alg1} are analyzed as follows.  The convergence of Algorithm \ref{alg1} is guaranteed since the alternating optimization of variables ${\bm \vartheta}$ and $\phi$ results in a non-decreasing capacity in each iteration, which is bounded below a finite value.  Moreover, the convergence behavior will be validated through simulations in Section \ref{sec4}.  As for computational complexity, from step 3 to 10, the complexities of calculating $\sum_{n = 1}^N {\bf P}_{1n} {\bf f}\left(\theta_n^{(i)}\right) {\bf f}\left(\theta_n^{(i)}\right)^H {\bf P}_{1n}^H$, $\phi^{(i)}$, $\left\{{\bf P}_{1n}^H {\bf g}(\phi) {\bf g}(\phi)^H {\bf P}_{1n}\right\}_{n=1}^N$, ${\bm \vartheta}^{(i)}$,  and $C_{\rm MISO}\left({\bm \vartheta}^{(i)}, \phi^{(i)}\right)$ are $\mathcal{O}(N)$, $\mathcal{O}(1)$, $\mathcal{O}(N)$, $\mathcal{O}(N)$, and $\mathcal{O}(N)$, respectively.  Thus, the total computational complexity of Algorithm \ref{alg1} is $\mathcal{O}(N\hat{I}_{\max})$, which is linear with respect to both $N$ and $\hat{I}_{\max}$.}

Next, we consider the case of $M>1$ and $N=1$ in the SIMO system.  Here, the transmit PFV can be expressed as ${\bf f}(\theta)$, where $\theta$ represents the phase shift of the transmit PRA.  The effective channel vector simplifies to ${\bf h}(\theta, {\bm \varphi}) = {\bf G}({\bm \varphi})^H {\bf P} {\bf f}(\theta) \in \mathbb{C}^M$.  The polarized channel matrix in \eqref{PCM} can be rewritten as
\begin{equation}
{\bf P} = \left[ {\bf P}_{11}^H, {\bf P}_{21}^H, \cdots, {\bf P}_{M1}^H \right]^H \in \mathbb{C}^{2M\times 2}.
\end{equation}
For the case of $N=1$, there is only one data stream for transmission, and thus, the optimal transmit covariance matrix is readily determined by the maximum transmit power as ${\bf Q}^\star = [P_{\rm t}]$.  As a result, the channel capacity of the SIMO system is given by
\begin{align} \label{simo_capacity}
	C_{\rm SIMO}(\theta, {\bm \varphi}) &= \log_2\det\left( {\bf I}_M + \frac{1}{\sigma^2} {\bf h} (\theta, {\bm \varphi}) {\bf Q}^\star {\bf h} (\theta, {\bm \varphi})^H \right) \nonumber \\
	&\mathop = \limits^{(b)} \log_2\left( 1 + \frac{P_{\rm t}}{\sigma^2} \left\| {\bf h} (\theta, {\bm \varphi}) \right\|^2_2\right).
\end{align}
The equality marked by $(b)$ holds because of the identity $\det({\bf I}_p + {\bf A}{\bf B}) = \det({\bf I}_q + {\bf B}{\bf A})$ for ${\bf A}\in \mathbb{C}^{p\times q}$ and ${\bf B}\in \mathbb{C}^{q\times p}$.  To maximize the channel capacity of the SIMO system, the optimization problem can be formulated as
\begin{subequations}
	\label{problem_simo}
	\begin{align}
		\max_{\theta, {\bm \varphi}} ~ & ~ \left\| {\bf h} (\theta, {\bm \varphi}) \right\|^2_2 \\
		{\rm s.t.} ~ & ~ \eqref{cons_phi}, \\
		& ~ \theta \in [0,2\pi].
	\end{align}
\end{subequations}
Problem \eqref{problem_simo} can be solved similarly to problem \eqref{problem_miso} (i.e., Algorithm \ref{alg1}) by replacing $\left\{ N, {\bm \vartheta}, \phi, \{{\bf P}_{1n}\}_{n=1}^{N} \right\}$ with $\left\{ M, \theta, {\bm \varphi}, \{{\bf P}_{m1}^H\}_{m=1}^{M} \right\}$, respectively.  Specifically, for given ${\bm \varphi}$, the optimal phase shift $\theta$ to maximize the objective function $\left\|{\bf h}(\theta)\right\|^2_2$ is given by
\begin{equation}
	\theta^\star = \angle{\left[ \sum_{m = 1}^M {\bf P}_{m1}^H {\bf g}(\phi_m) {\bf g}(\phi_m)^H {\bf P}_{m1} \right]_{21}}.
\end{equation} 
For given $\theta$, the optimal phase shift $\phi_m$ to maximize the objective function $\left\|{\bf h}({\bm \varphi})\right\|^2_2$ is similarly given by
\begin{equation} \label{opt_phi_simo}
	\phi_m^\star = \angle{\left[{\bf P}_{m1} {\bf f}(\theta) {\bf f}(\theta)^H {\bf P}_{m1}^H\right]_{21}},\;m = 1,2,\cdots,M.
\end{equation}
The phase shifts given in \eqref{opt_phi_simo} represent the optimal solution for polarforming applied at the receiver with an FPA at the transmitter, referred to as receive polarforming, which will be further analyzed in Section \ref{sec4}.  It is worth noting that even for the simplest case of the SISO system with $M=N=1$, finding a global optimum for problem \eqref{problem_miso} is still challenging.  In this paper, the solution for the SISO system can be viewed as a special case of the MISO/SIMO system.

\subsection{Capacity Maximization for MIMO System} \label{sec3-3}
Next, we develop a capacity maximization method by maximizing the upper bound of the MIMO channel capacity, which simplifies problem \eqref{opt_problem} by reducing the variables to only the transmit and receive PSVs, ${\bm \vartheta}$ and ${\bm \varphi}$.  By leveraging this simplification, we can apply an alternating optimization approach to solve the problem efficiently.

Since we assume full CSI is available at the transmitter and receiver, we can employ singular value decomposition (SVD)-based transmit precoding and receive combining to determine the MIMO channel capacity in \eqref{capacity_mimo}.  Specifically, by applying the truncated SVD, we decompose ${\bf H}({\bm \vartheta}, {\bm \varphi}) = {\bf U}{\bf \Lambda} {\bf V}^H$, where ${\bf U}\in\mathbb{C}^{M\times S}$, ${\bf V}\in\mathbb{C}^{N\times S}$, and ${\bf \Lambda}\in\mathbb{C}^{S\times S}$, with $S$ denoting the rank of ${\bf H}({\bm \vartheta}, {\bm \varphi})$.  The optimal covariance matrix ${\bf Q}$ is then given by
\begin{equation} \label{opt_Q}
	{\bf Q}^\star = {\bf V} {\rm diag}\left\{ \sqrt{p_1^\star}, \sqrt{p_2^\star}, \cdots, \sqrt{p_S^\star}  \right\} {\bf V}^H,
\end{equation}
where $p_s^\star$ is the optimal power allocated to the $s$-th data stream for $s = 1,2,\cdots,S$.  Based on the water-filling principle, $p_s^\star = \max\left( 0, \frac{1}{p_0} - \frac{\sigma^2}{\lambda_s^2} \right)$ subject to the total power constraint $\sum\nolimits_{s=1}^{S} p_s^\star = P_{\rm t}$, where $p_0$ is referred to as the water-filling level, and $\lambda_s$ is the $s$-th singular value of ${\bf H}({\bm \vartheta}, {\bm \varphi})$.   The optimal transmit precoding matrix and receive combining matrix for achieving the maximum capacity are ${\bf V} {\rm diag}\left\{ \sqrt{p^\star_1}, \sqrt{p^\star_2}, \cdots, \sqrt{p^\star_S} \right\}$ and ${\bf U}$, respectively.  Thus, the MIMO channel capacity can be expressed as
\begin{align} \label{waterfilling_capacity}
	C({\bm \vartheta}, {\bm \varphi}) = \sum_{s=1}^S\log_2\left( 1 + \frac{p_s^\star }{\sigma^2} \lambda_s^2 \right).
\end{align}

Based on \eqref{waterfilling_capacity}, we propose the following theorem to derive an upper bound of the MIMO channel capacity.

\begin{theorem} \label{theorem2}
The upper bound of the MIMO channel capacity in \eqref{waterfilling_capacity} is given by
\begin{equation} \label{upper_bound}
	\bar{C}({\bm \vartheta}, {\bm \varphi}) = S \log_2 \left( \frac{1}{Sp_0\sigma^2} {\rm Tr}\left( {\bf H}({\bm \vartheta}, {\bm \varphi}) {\bf H}({\bm \vartheta}, {\bm \varphi})^H \right) + \frac{\bar{S}}{S}	\right),
\end{equation}
where $\bar{S}$ denotes the number of data streams allocated zero power according to the water-filling principle.
\end{theorem}

\begin{IEEEproof}
Let $\bar{\mathcal S}$ denote the set of indices corresponding to data streams allocated zero power.  Define $\bar{p}_s^\star \triangleq \frac{1}{p_0} - \frac{\sigma^2}{\lambda_s^2}$.  Then, the channel capacity in \eqref{waterfilling_capacity} is upper-bounded by
\begin{align} \label{derive_upper_bound}
	C({\bm \vartheta}, {\bm \varphi}) &\mathop {\le}\limits^{(c)} S \log_2\left(1 + \frac{1}{S} \sum_{s = 1}^{S} \frac{p_s^\star}{\sigma^2} \lambda_s^2 \right) \nonumber \\
	& = S\log_2 \left( 1 + \frac{1}{S} \sum_{s=1}^S \frac{\bar{p}_s^\star}{\sigma^2}\lambda_s^2 - \frac{1}{S} \sum_{s\in\bar{\mathcal{S}}} \frac{\bar{p}_s^\star}{\sigma^2}\lambda_s^2 \right) \nonumber \\
	&  = S\log_2 \left( \frac{1}{Sp_0\sigma^2} \left( \sum_{s=1}^S \lambda_s^2 - \sum_{s\in\bar{\mathcal{S}}} \lambda_s^2 \right) + \frac{\bar{S}}{S} \right) \nonumber \\
	&  \mathop {\le}\limits^{(d)}  S \log_2 \left( \frac{1}{Sp_0\sigma^2} \sum_{s=1}^S \lambda_s^2 + \frac{\bar{S}}{S}	\right).
\end{align}
The inequality marked by $(c)$ is justified by the Jensen's inequality \cite{ref_TV05}.  The equality marked by $(d)$ holds when all data streams are allocated non-zero power, i.e., when $\bar{S} = 0$.  A channel is allocated zero power when its associated singular value ${\lambda}_s$ is sufficiently small, such that $\bar{p}_s^\star = \frac{1}{p_0} - \frac{\sigma^2}{\lambda_s^2} \le 0$.  Therefore, it is reasonable to omit the term $\sum_{s\in\bar{\mathcal{S}}} \lambda_s^2$ to establish the inequality\footnote{It is worth noting that the capacity upper bound given in \eqref{upper_bound} holds across the entire range of SNR, which is different from the high-SNR capacity approximation used in \cite{ref_KM15}.} marked by $(d)$.

As the sum of the eigenvalues of a matrix equals its trace, we have
\begin{equation} \label{lambda_trace}
	\sum_{s=1}^S \lambda_s^2 = {\rm Tr}\left( {\bf H}({\bm \vartheta}, {\bm \varphi}) {\bf H}({\bm \vartheta}, {\bm \varphi})^H \right).
\end{equation}
By substituting \eqref{lambda_trace} into \eqref{derive_upper_bound}, we obtain the upper bound\footnote{ The upper bound is tight at high SNR, when the inequality $(d)$ in \eqref{derive_upper_bound} holds with equality approximately.  At low SNR, however, tightness is compromised due to the term $\sum_{s\in\bar{\mathcal{S}}} \lambda_s^2$, which is omitted in \eqref{derive_upper_bound}.} of the channel capacity in \eqref{upper_bound}, and the proof of the theorem is complete.
\end{IEEEproof}

According to Theorem \ref{theorem2}, by maximizing the upper bound of the channel capacity given in \eqref{upper_bound}, we simplify problem \eqref{opt_problem} into the following optimization problem
\begin{subequations}
	\label{problem_mimo}
	\begin{align}
		\max_{ {\bm \vartheta}, {\bm \varphi}} ~ & ~ {\rm Tr}\left( {\bf H}({\bm \vartheta}, {\bm \varphi}) {\bf H}({\bm \vartheta}, {\bm \varphi})^H \right) \\
		{\rm s.t.} ~ & ~ \eqref{cons_theta},\eqref{cons_phi}.
	\end{align}
\end{subequations}
This problem remains challenging to solve directly due to the exponential terms in the PFVs.  To address this, we adopt an alternating optimization approach, which decomposes the problem into more tractable subproblems.  The details of these subproblems are presented in the following.

First, we consider the subproblem for optimizing the $m$-th receive PRA's phase shift, while the phase shifts of the other transmit and receive PRAs remain fixed.  In other words, we aim to optimize $\phi_m$ with given $\{\phi_\iota,\iota\ne m\}_{\iota=1}^M$ and ${\bm \vartheta}$.  Recalling that ${\bf H}({\bm \vartheta}, {\bm \varphi}) = {\bf G}({\bm \varphi})^H {\bf P} {\bf F}({\bm \vartheta})$, the entry in the $m$-th row and $n$-th column of the channel matrix ${\bf H}({\bm \vartheta}, {\bm \varphi})$ can be expressed as $h(\theta_n, \phi_m) = {\bf g}(\phi_m)^H {\bf P}_{mn} {\bf f}(\theta_n)$.  Based on this, we can rewrite the objective function in problem \eqref{problem_mimo} as
\begin{align} \label{Am}
	{\rm Tr}&\left( {\bf H}({\bm \vartheta}, {\bm \varphi}) {\bf H}({\bm \vartheta}, {\bm \varphi})^H \right) \nonumber\\
	&= \sum_{m=1}^M\sum_{n=1}^N \left| h(\theta_n, \phi_m) \right|^2 \nonumber\\
	&= \sum_{m=1}^M {\bf g}(\phi_m)^H \left( \sum_{n=1}^N {\bf P}_{mn} {\bf f}(\theta_n) {\bf f}(\theta_n)^H {\bf P}_{mn}^H \right) {\bf g}(\phi_m) \nonumber\\
	&= {\bf g}(\phi_m)^H {\bf A}_m {\bf g}(\phi_m) + \sum_{\iota=1,\iota\ne m}^M {\bf g}(\phi_\iota)^H {\bf A}_\iota {\bf g}(\phi_\iota),
\end{align}
where ${\bf A}_m \triangleq \sum_{n=1}^N {\bf P}_{mn} {\bf f}(\theta_n) {\bf f}(\theta_n)^H {\bf P}_{mn}^H$ for all $m = 1,2,\cdots,M$.  For given $\{\phi_\iota,\iota\ne m\}_{\iota=1}^M$ and ${\bm \vartheta}$, the term $\sum_{\iota=1,\iota\ne m}^M {\bf g}(\phi_\iota)^H {\bf A}_\iota {\bf g}(\phi_\iota)$ is constant.  Thus, problem \eqref{problem_mimo} simplifies to maximizing ${\bf g}(\phi_m)^H {\bf A}_m {\bf g}(\phi_m)$ subject to the phase shift constraint $\phi_m \in [0, 2\pi]$.  It can be observed that the matrix ${\bf A}_m$ is Hermitian, as it is a sum of Hermitian matrices.  According to Theorem\;\ref{theorem1}, the optimal phase shift $\phi_m$ for the $m$-th receive PRA in this subproblem is given by
\begin{equation} \label{opt_phim}
	\phi_m^\star = \angle{[{\bf A}_m]_{21}}.
\end{equation}

Next, we consider the subproblem for optimizing the $n$-th transmit PRA's phase shift, while the phase shifts of the other transmit and receive PRAs are fixed.  In this subproblem, our goal is to optimize $\theta_n$ with given $\{\theta_\ell, \ell\ne n\}_{\ell=1}^N$ and ${\bm \varphi}$.  Similar to the procedure for optimizing $\phi_m$, we define
\begin{equation} \label{Bn}
	{\bf B}_n \triangleq \sum_{m=1}^M {\bf P}_{mn}^H {\bf g}(\phi_m) {\bf g}(\phi_m)^H {\bf P}_{mn}.
\end{equation}
Then, we can rewrite the objective function in problem \eqref{problem_mimo} as
\begin{align}
	{\rm Tr}&\left( {\bf H}({\bm \vartheta}, {\bm \varphi}) {\bf H}({\bm \vartheta}, {\bm \varphi})^H \right) \nonumber\\
	&= {\bf f}(\theta_n)^H {\bf B}_n {\bf f}(\theta_n) + \sum_{\ell=1,\ell\ne n}^N {\bf f}(\theta_\ell)^H {\bf B}_\ell {\bf f}(\theta_\ell).
\end{align}
Since the term $\sum_{\ell=1,\ell\ne n}^N {\bf f}(\theta_\ell)^H {\bf B}_\ell {\bf f}(\theta_\ell)$ remains constant for given $\{\theta_\ell, \ell\ne n\}_{\ell=1}^N$ and ${\bm \varphi}$, the subproblem reduces to finding the optimal phase shift $\theta_n$ that maximizes ${\bf f}(\theta_n)^H {\bf B}_n {\bf f}(\theta_n)$ subject to $\theta_n\in[0,2\pi]$.  Given that the matrix ${\bf B}_n$ is Hermitian, the optimal phase shift $\theta_n$ for the $n$-th transmit PRA in this subproblem is given by
\begin{equation} \label{opt_thetan}
	\theta_n^\star = \angle{[{\bf B}_n]_{21}}.
\end{equation}

\begin{algorithm}[!t]
	\caption{Proposed Solution for Solving Problem \eqref{opt_problem}}
	\label{alg2}
	\footnotesize
	\renewcommand{\algorithmicrequire}{\textbf{Input:}}
	\renewcommand{\algorithmicensure}{\textbf{Output:}}
	\begin{algorithmic}[1]
		\REQUIRE $M$, $N$, $P_{\rm t}$, $\sigma$, ${\bf P}$, $\epsilon_2$.
		\ENSURE ${\bm \vartheta}$, ${\bm \varphi}$, ${\bf Q}$.
		\STATE Initialize ${\bm \vartheta}^{(0)} = {\bf 0}_{N}$ and ${\bm \varphi}^{(0)} = {\bf 0}_{M}$.
		\FOR{$i=1\rightarrow\tilde{I}_{\max}$}
		\STATE Update ${\bm \vartheta}^{(i)} \leftarrow {\bm \vartheta}^{(i-1)}$ and ${\bm \varphi}^{(i)} \leftarrow {\bm \varphi}^{(i-1)}$.
		\FOR{$m=1\rightarrow M$}
		\STATE Given $\left\{\phi_\iota^{(i)},\iota\ne m\right\}_{\iota=1}^M$ and ${\bm \vartheta}^{(i)}$, calculate ${\bf A}_m$ via \eqref{Am}.
		\STATE Update $\phi_m^{(i)}$ according to \eqref{opt_phim}.
		\ENDFOR
		\FOR{$n=1\rightarrow N$}
		\STATE Given $\left\{\theta_\ell^{(i)},\ell\ne n\right\}_{\ell=1}^N$ and ${\bm \varphi}^{(i)}$, calculate ${\bf B}_n$ via \eqref{Bn}.
		\STATE Update $\theta_n^{(i)}$ according to \eqref{opt_thetan}.
		\ENDFOR
		\STATE Calculate the candidate ${\bf Q}^{(i)}$ in $\eqref{opt_Q}$ using ${\bm \vartheta}^{(i)}$ and ${\bm \varphi}^{(i)}$.
		\IF{Increase of the channel capacity in \eqref{waterfilling_capacity} is below $\epsilon_2$}
		\STATE Set ${\bm \vartheta} = {\bm \vartheta}^{(i-1)}$.
		\STATE Set ${\bm \varphi} = {\bm \varphi}^{(i-1)}$.
		\STATE Set ${\bf Q} = {\bf Q}^{(i-1)}$.
		\STATE Break.
		\ENDIF
		\STATE Set ${\bm \vartheta} = {\bm \vartheta}^{(i)}$.
		\STATE Set ${\bm \varphi} = {\bm \varphi}^{(i)}$.
		\STATE Set ${\bf Q} = {\bf Q}^{(i)}$.
		\ENDFOR
		\RETURN ${\bm \vartheta}$, ${\bm \varphi}$, ${\bf Q}$.
	\end{algorithmic}
\end{algorithm}

Algorithm \ref{alg2} iteratively solves problem \eqref{problem_mimo} by alternating between optimizing the phase shifts of the transmit and receive PRAs.  The algorithm initializes the transmit and receive PSVs to linear polarization, i.e., ${\bm \vartheta}^{(0)}={\bf 0}_N$ and ${\bm \varphi}^{(0)} = {\bf 0}_M$, and proceeds with an iterative process.  In each iteration, the transmit and receive PSVs are sequentially updated using the candidate phase shifts derived in \eqref{opt_phim} and \eqref{opt_thetan}.  The covariance matrix ${\bf Q}^{(i)}$ is computed based on the candidate PSVs to evaluate the increase in channel capacity.  If the increase in capacity falls below a predefined threshold $\epsilon_2$ or the maximum number of iterations $\tilde{I}_{\max}$ is reached, the process terminates.  It is worth noting that although the candidate ${\bm \vartheta}^{(i)}$ and ${\bm \varphi}^{(i)}$ in each iteration can increase the upper bound of the channel capacity in \eqref{upper_bound}, the procedure is not guaranteed to reach the global optimum for problem \eqref{opt_problem}. { The proposed alternating optimization procedure is deterministic, with each iteration depending solely on the global channel state (i.e., the polarized channel matrix ${\bf P}$) and the fixed system configuration (e.g., the antenna numbers $M$ and $N$).  Therefore, the transmitter and receiver do not need to exchange intermediate information in every iteration, and both sides can independently execute the algorithm and obtain the final candidate solution.}

{ The convergence and computational complexity of Algorithm \ref{alg2} are analyzed as follows.  The convergence of Algorithm \ref{alg2} is guaranteed because the alternating optimization of the variables ${\bm \vartheta}$ and ${\bm \varphi}$ generates a non-decreasing sequence of objective function values in problem \eqref{problem_mimo}.  Besides, the convergence behavior will be further validated through simulations in Section \ref{sec4}.  In terms of computational complexity, the steps from 3 to 21 involve the following operations, i.e., calculating $\{{\bf A}_m\}_{m = 1}^M$, $\{\phi_m^{(i)}\}_{m=1}^M$, $\{{\bf B}_n\}_{n = 1}^N$, $\{\theta_n^{(i)}\}_{n=1}^N$, ${\bf Q}^{(i)}$, and $C\left({\bm \vartheta}^{(i)}, {\bm \varphi}^{(i)}\right)$, corresponding to the complexities of $\mathcal{O}(MN)$, $\mathcal{O}(M)$, $\mathcal{O}(MN)$, $\mathcal{O}(N)$, $\mathcal{O}(MN\min(M,N))$, and $\mathcal{O}(S\log_2S)$, respectively.  The total computational complexity of Algorithm\;\ref{alg2} is therefore $\mathcal{O}(MN\tilde{I}_{\max}\min(M,N) + MN\tilde{I}_{\max} + S\tilde{I}_{\max}\log_2S)$.}

\subsection{Practical Considerations} \label{sec3-4}
In this subsection, we discuss the practical aspects of polarforming that may influence its effectiveness.

\subsubsection{Antenna XPI}
Ideally, a polarized antenna generates an oscillating electric field only along its designated orientation with no components leaking into the orthogonal orientation.  However, in practical antenna design, cross-polar leakage inevitably occurs due to imperfections in the antenna architecture and design.  Such leakage is well-known in antenna theory \cite{ref_Balanis16} and commonly characterized using the antenna XPI.  Analytically, the presence of leakages can contribute to depolarization effects in both LoS and NLoS channels.  To account for this impact, the polarized channel matrix ${\bf P}_{mn}$ in \eqref{PCM} can be modeled as \cite{ref_OCG08}
\begin{equation} \label{P_XPI}
	\mathbf{P}^{\rm XPI}_{mn} = \mathbf{X}_\mathrm{r}^{1/2} \mathbf{P}_{mn} \mathbf{X}_\mathrm{t}^{1/2}.
\end{equation}
The coupling matrices of the transmit and receive antennas with normalized channel power are expressed as
\begin{align}
	\mathbf{X}_\mathrm{t} &= \frac{1}{\sqrt{\mu_{\rm t}+1}} \begin{bmatrix}
		{1}&{\sqrt{\mu_{\rm t}}}\\
		{\sqrt{\mu_{\rm t}}}&{1}
	\end{bmatrix},\\
	\mathbf{X}_\mathrm{r} &= \frac{1}{\sqrt{\mu_{\rm r}+1}} \begin{bmatrix}
		{1}&{\sqrt{\mu_{\rm r}}}\\
		{\sqrt{\mu_{\rm r}}}&{1}
	\end{bmatrix},
\end{align}
where $\mu_\mathrm{t}$ and $\mu_\mathrm{r}$ stand for the inverse XPI of the transmit and receive antennas, respectively.  In practice, the values of $\mu_\mathrm{t}$ and $\mu_\mathrm{r}$ do not exceed one.  Note that imperfect antenna XPI not only alters channel depolarization but also degrades the correlation of the polarized channel.

\subsubsection{Polarized Channel Correlation}
In wireless communication systems, when the scattering is insufficient to decorrelate the channels, the elements of ${\bf H}({\bm \vartheta}, {\bm \varphi})$ become correlated, known as spatial correlation.  Insufficient reflections, diffractions, and scattering of EM waves can also lead to the correlation of polarized channels.  This correlation may deteriorate the performance of a polarized system by reducing its ability to effectively separate signals.  By denoting $\widetilde{\bf P}_{mn}$ as the NLoS component of ${\bf P}_{mn}$, the correlated channel matrix of normalized channel power is given by \cite{ref_Coldrey08}
\begin{equation}
	\widetilde{\bf P}_{mn}^{\rm Corr} = \mathbf{C}_{\rm r}^{1/2} \widetilde{\bf P}_{mn} {\bf C}_{\rm t}^{1/2},
\end{equation}
where $\mathbf{C}_{\rm t}$ and $\mathbf{C}_{\rm r}$ are the correlation matrices, defined as
\begin{align}
	{\bf C}_{\rm t} &= \frac{1}{\sqrt{|\nu_{\rm t}|^2 + 1}}\begin{bmatrix}
		{1}&{\nu_{\rm t}^*}\\
		{\nu_{\rm t}}&{1}
	\end{bmatrix},\\
	{\bf C}_{\rm r} &= \frac{1}{\sqrt{|\nu_{\rm r}|^2 + 1}}\begin{bmatrix}
		{1}&{\nu_{\rm r}^*}\\
		{\nu_{\rm r}}&{1}
	\end{bmatrix}.
\end{align}
Here, $\nu_{\rm t}$ and $\nu_{\rm r}$ represent the complex-valued correlation coefficients at the transmitter and receiver, respectively.

\subsubsection{Polarized Channel Estimation}
CSI is essential for optimizing the system performance, as its accuracy directly influences the effectiveness of polarforming.  Precise CSI allows the system to better adapt to channel variations and mitigate channel depolarization, as well as to further exploit polarization diversity.  In conventional DPA systems, each DPA is connected to two dedicated RF chains, and the CSI can be attained using standard channel estimation techniques, such as least squares (LS), minimum mean square error (MMSE), and maximum likelihood (ML) estimation \cite{ref_TV05}.  In contrast, each PRA of the considered system employs only a single RF chain, which poses a challenge for estimating the CSI of a two-by-two polarized channel matrix.  Thus, it will be interesting to devise new methods to acquire the CSI of polarized channels more efficiently in future work.
\begin{figure}[!t]
	\centering
	\subfigure[Algorithm \ref{alg1}]{
		\centering
		\includegraphics[width=0.45\columnwidth]{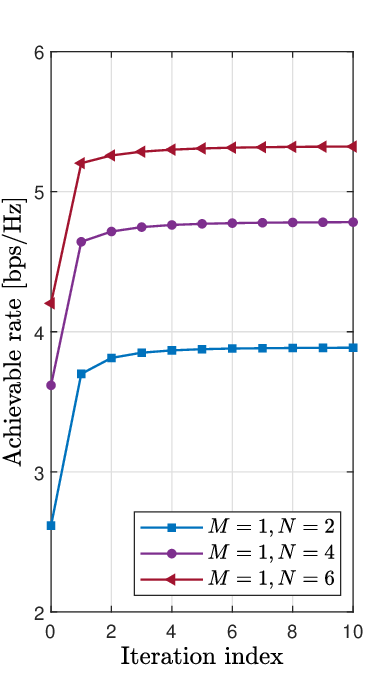}}
	\subfigure[Algorithm \ref{alg2}]{
		\centering
		\includegraphics[width=0.45\columnwidth]{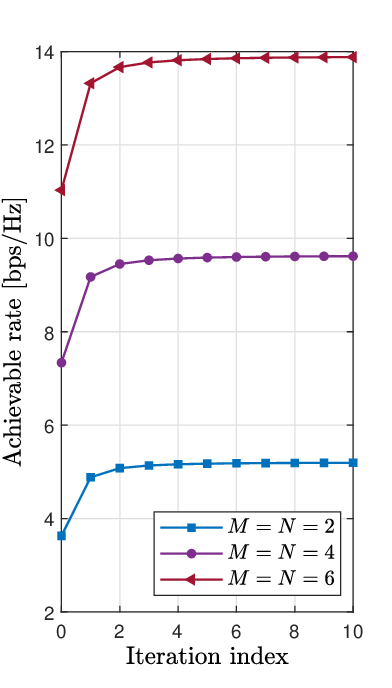}}
	\caption{{ Convergence behavior of the proposed algorithms.}}
	\label{fig_conv}
\end{figure}

\section{Simulation Results} \label{sec4}
We carry out extensive simulations and present their results in this section to evaluate the performance gains of the proposed system over conventional/benchmark systems and validate the effectiveness of the proposed algorithms.

{ In the simulations, we assume that the PSs for polarforming can be flexibly adjusted within the range of $0$ to $2\pi$.  The depolarization of the LoS channel is very complicated and depends on the antenna orientation \cite{ref_CP12}.  Therefore, we focus on the Rayleigh fading channel, which is also considered in \cite{ref_KM15}.  The polarized channel matrix in \eqref{PCM} is then modeled as ${\bf P}_{mn} = {\bf \Psi}\odot{\bf H}^{\rm i.i.d.}_{mn}$ \cite{ref_HCS16}, where
\begin{equation}
	{\bf \Psi} = \frac{1}{\sqrt{\chi+1}} \begin{bmatrix}1&\sqrt{\chi}\\\sqrt{\chi}&1\end{bmatrix},
\end{equation}
and $\chi$ represents the inverse XPD indicating the degree of channel depolarization.  Unless otherwise stated, the inverse XPD will be set to $\chi = 0.2$ as in \cite{ref_CH24}.  The elements of the matrix ${\bf H}^{\rm i.i.d.}_{mn} \in \mathbb{C}^{2\times 2}$ are independent and identically distributed (i.i.d.) and CSCG-distributed random variables with equal variance of one after normalization.  Due to the normalization of the channel matrix, the average SNR of the proposed system is only determined by the transmit power $P_{\rm t}$ and the noise power $\sigma^2$ at the receiver, i.e., $\text{SNR} = \frac{P_{\rm t}}{\sigma^2}$.  Furthermore, the convergence thresholds for the relative increment of the objective function are set to $\epsilon_1 = \epsilon_2 = 10^{-3}$ for both Algorithms \ref{alg1} and \ref{alg2}.  The maximum number of iterations is limited to $\hat{I}_{\max} = \tilde{I}_{\max} = 20$ to ensure efficient convergence within a reasonable computational cost.  To ensure the robustness of the simulations, the results are obtained by averaging over $10^4$ independent Monte Carlo channel realizations.}

{ First of all, the convergence behavior of the proposed algorithms for different values of $M$ and $N$ is illustrated in Fig.\;\ref{fig_conv}, with $\text{SNR} = 5$ dB.  The results demonstrate that the achievable rate consistently increases and reaches a maximum value after six iterations for all values of $M$ and $N$, which validates the convergence analysis in Sections \ref{sec3-2} and \ref{sec3-3}.  Moreover, as an example, the final converged achievable rate improves by 44.3\% over the initial value for $M=1,N=2$ and by 39.9\% for $M = N = 2$.}

\begin{table}[!t]
	\centering
	\renewcommand{\arraystretch}{1.25}
	\caption{Average CPU Time per Channel Realization (ms)}
	\label{tab2}
	\begin{tabular}{|c|c|c|c|c|c|c|}
		\hline
		{\bf $(M,N)$} & {\bf Proposed} & {\bf SPRA} & {\bf PAA} & {\bf CPA} & {\bf LPA} & {\bf Exhaustive} \\
		\hline\hline
		$(1,2)$ & $0.24$ & $0.09$ & $0.19$ & $0.02$ & $0.02$ & $2.0\times 10^{2}$ \\
		\hline
		$(2,2)$ & $0.39$ & $0.14$ & $0.27$ & $0.02$ & $0.02$ & $5.4\times 10^{3}$ \\
		\hline
		$(4,4)$ & $0.70$ & $0.28$ & $0.55$ & $0.03$ & $0.03$ & --- \\
		\hline
	\end{tabular}
\end{table}

To complement the complexity analysis in Sections \ref{sec3-2} and \ref{sec3-3}, Table \ref{tab2} reports the average CPU time per channel realization in milliseconds (ms) for the proposed algorithms, the benchmark schemes, and the exhaustive search with a $20$-point grid per PS phase, measured in MATLAB R2024a on a single core of an AMD Ryzen 9 7950X processor.  The proposed alternating optimization completes within $1$ ms, which is comparable to the benchmark schemes and over two orders of magnitude faster than the exhaustive search.  For $(M,N)=(4,4)$, the exhaustive search becomes computationally prohibitive, since its search space grows to $20^{M+N}$ grid points with an estimated runtime exceeding one week per channel realization.  Nevertheless, the proposed alternating optimization attains on average over $99\%$ and about $98\%$ of the exhaustive-search rate for $(M,N)=(1,2)$ and $(M,N)=(2,2)$, respectively, which confirms that polarforming achieves near-optimal performance at a very low computational cost.

\begin{figure}[!t]
	\centering
	\includegraphics[width=0.9\linewidth]{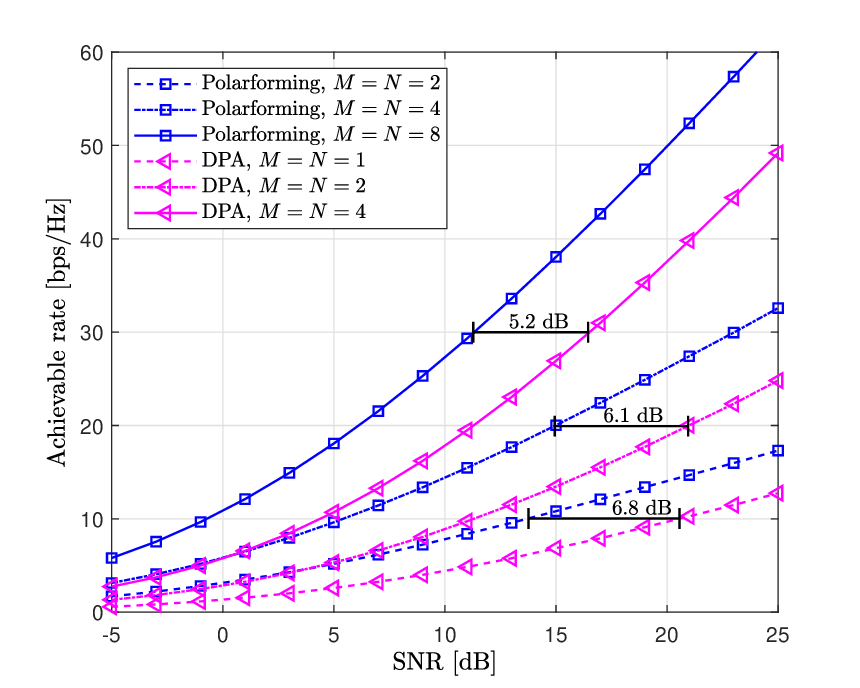}
	\caption{{ Comparison of polarforming with the DPA scheme.}}
	\label{fig_compare_DPA}
\end{figure}

\begin{figure*}[!t]
	\centering
	\subfigure[Transmit polarforming with a single-LPA at the receiver]{ \label{fig_txp_LPA}
		\centering
		\includegraphics[width=0.9\columnwidth]{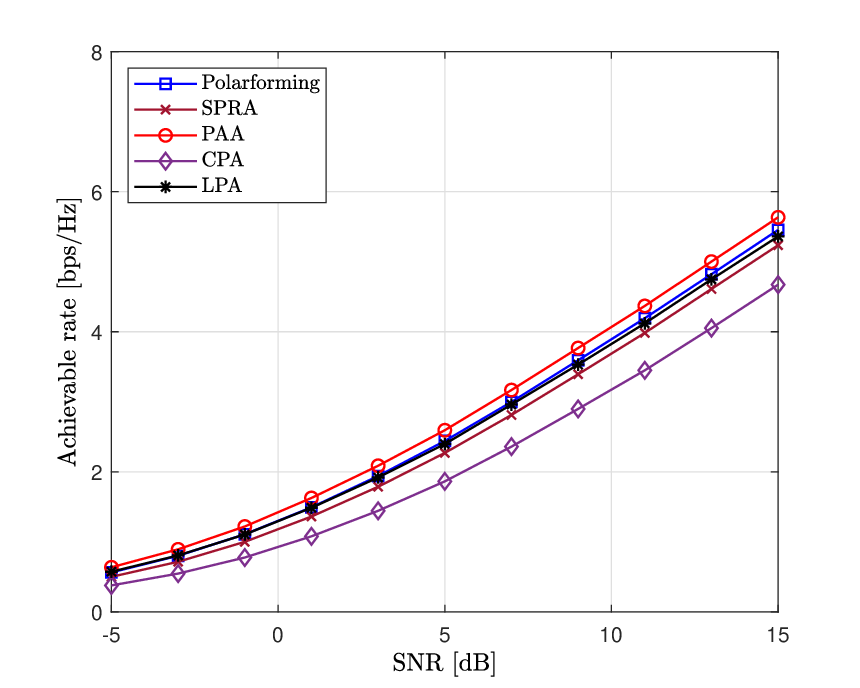}}
	\subfigure[Receive polarforming with a single-LPA at the transmitter]{ \label{fig_rxp_LPA}
		\centering
		\includegraphics[width=0.9\columnwidth]{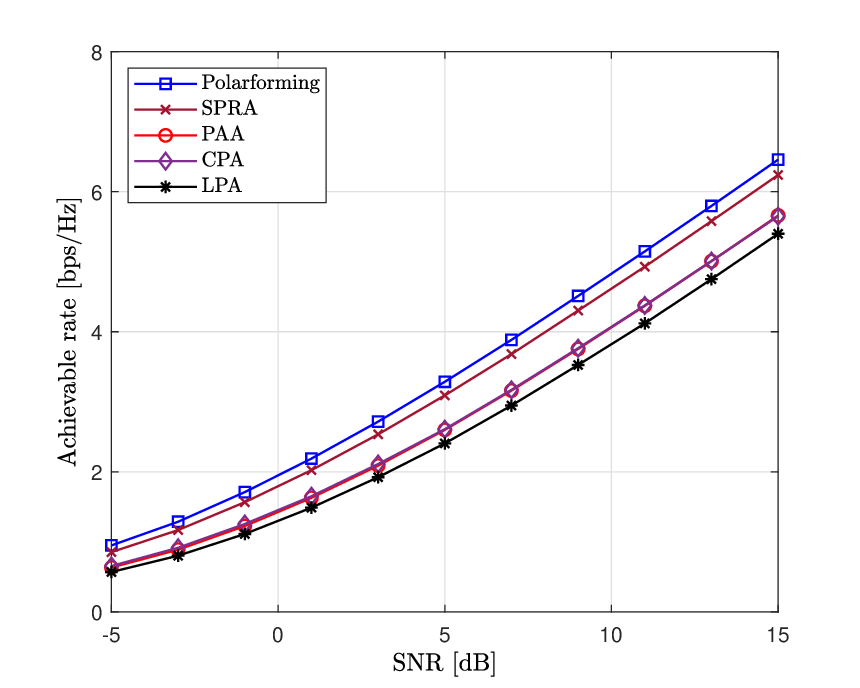}}
	\subfigure[Transmit polarforming with a single-CPA at the receiver]{  \label{fig_txp_CPA}
		\centering
		\includegraphics[width=0.9\columnwidth]{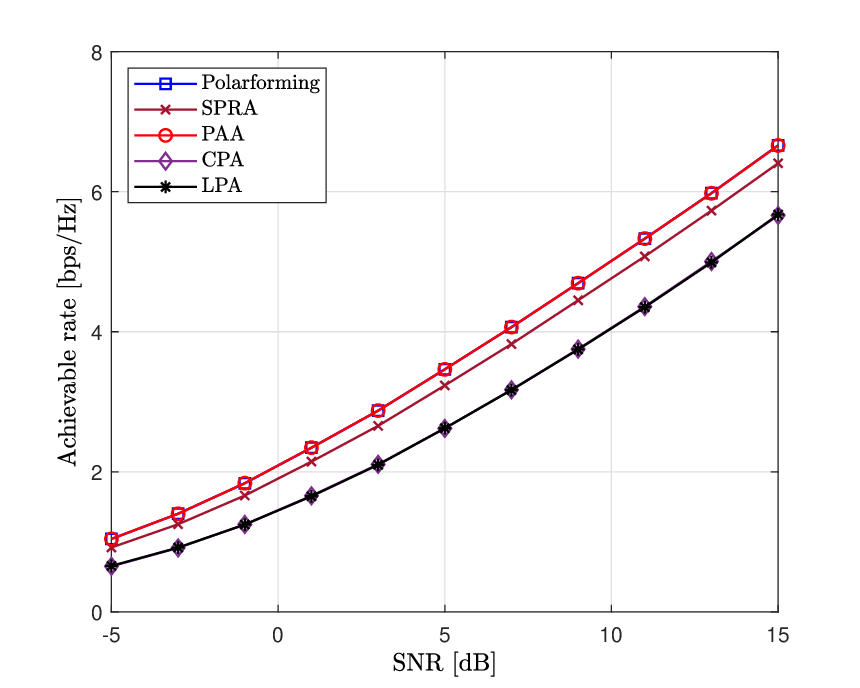}}
	\subfigure[Receive polarforming with a single-CPA at the transmitter]{  \label{fig_rxp_CPA}
		\centering
		\includegraphics[width=0.9\columnwidth]{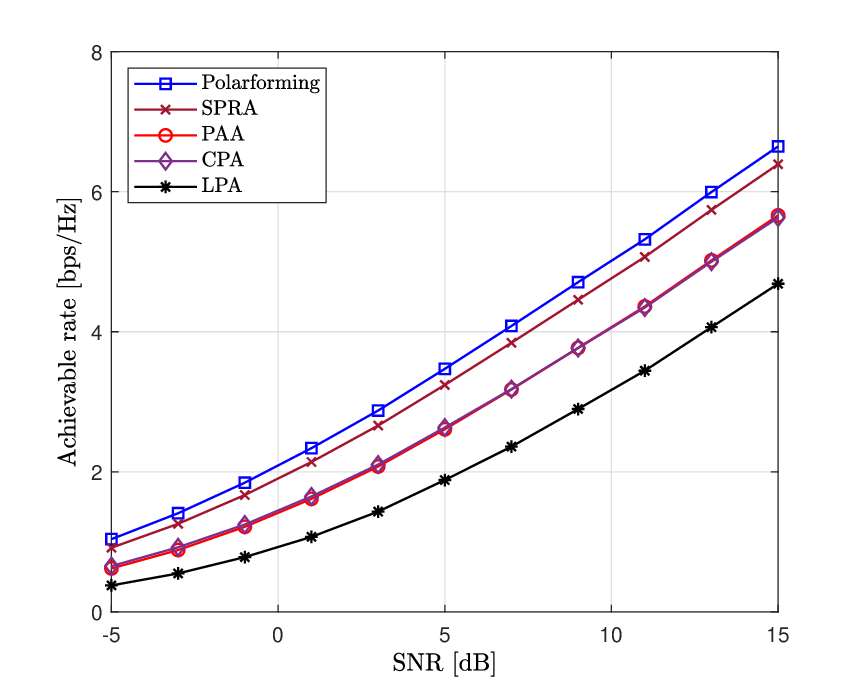}}
	\caption{Achievable rate of transmit/receive polarforming versus SNR in the MISO/SIMO system when it is applied at the transmitter/receiver only.}
	\label{fig_PF}
\end{figure*}

\begin{figure*}[!t]
	\centering
	\subfigure[SISO ($M=N=1$)]{
		\centering
		\includegraphics[width=0.9\columnwidth]{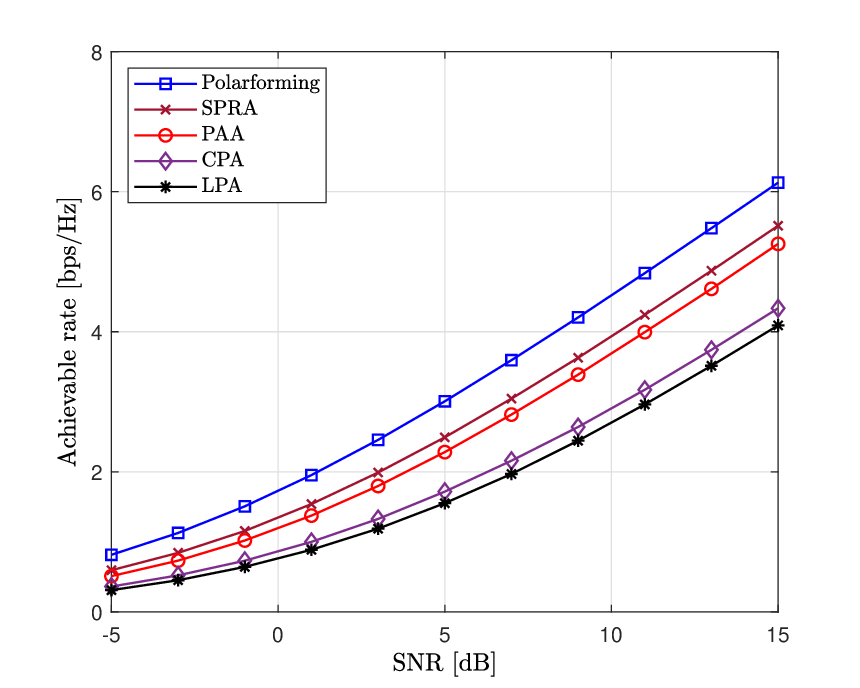}}
	\subfigure[MISO ($M=1,N=2$)]{
		\centering
		\includegraphics[width=0.9\columnwidth]{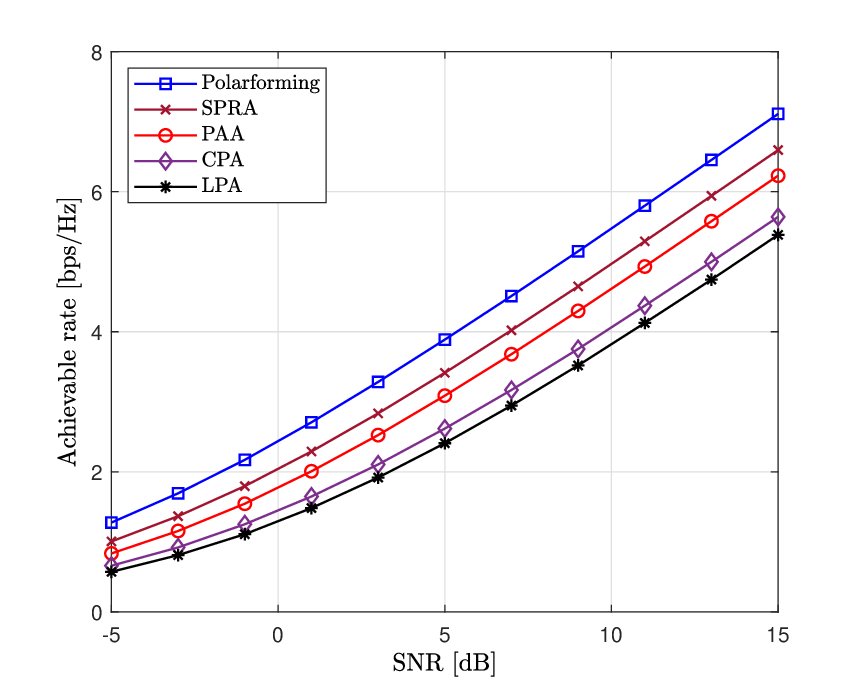}}
	\subfigure[SIMO ($M=2,N=1$)]{
		\centering
		\includegraphics[width=0.9\columnwidth]{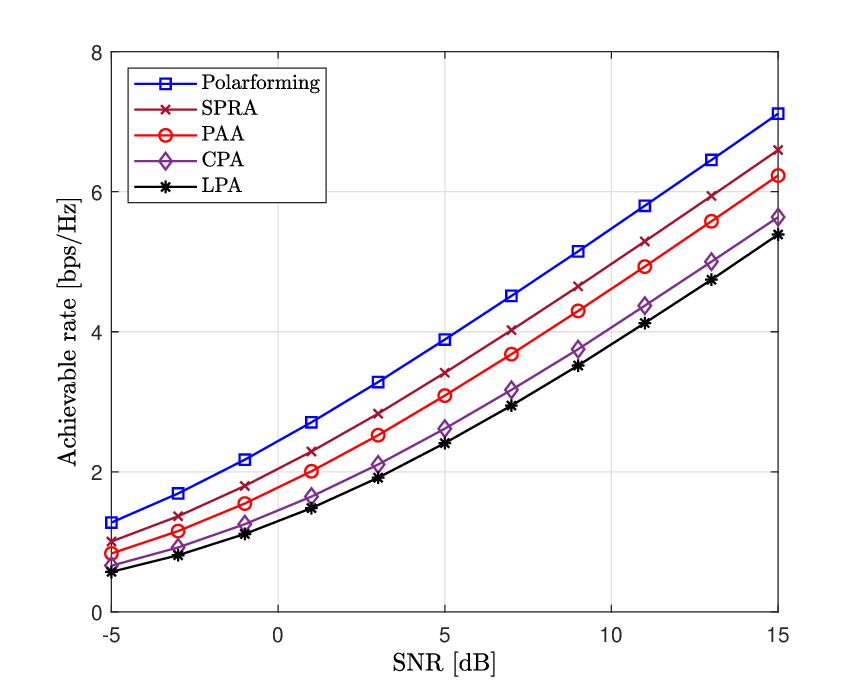}}
	\subfigure[MIMO ($M=N=2$)]{
		\centering
		\includegraphics[width=0.9\columnwidth]{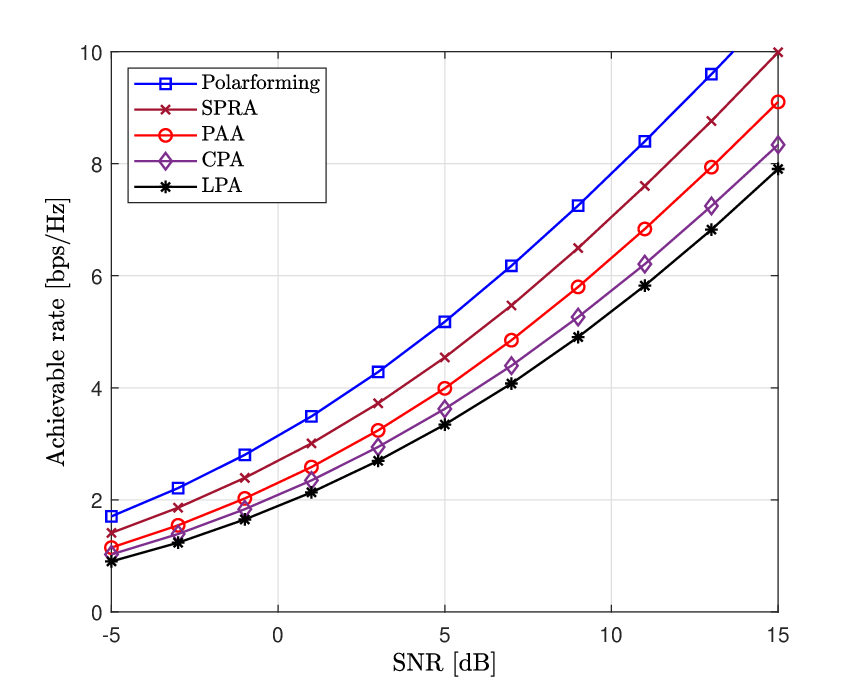}}
	\caption{Achievable rate of polarforming versus SNR when it is applied at both the transmitter and receiver.}
	\label{fig_rate_snr}
\end{figure*}

In comparison to the proposed scheme marked by ``Polarforming'' with PS-based PRAs, five benchmark schemes are considered, defined below.
\begin{itemize}
	\item \textbf{DPA}: This system comprises $N$ transmit and $M$ receive DPAs, and it can transmit up to $2N$ data streams as each DPA is equipped with two dedicated RF chains.  Although the channel matrix in \eqref{PCM} is inherently polarized, it is treated as a $2M\times 2N$ unpolarized channel.
	
	\item \textbf{SPRA}: This system consists of $N$ transmit and $M$ receive SPRAs, with each SPRA connected to a single RF chain.  According to \cite{ref_KRR15}, each antenna can switch among two polarization states, i.e., left-handed circular polarization and right-handed circular polarization.  In this scheme, the transmit and receive polarization vectors\footnote{A polarization vector is defined to represent the polarization of antennas in the form of the Jones vector \cite{ref_KM15}.  For PS-based PRAs, this polarization vector is referred to as the PFV, as defined in \eqref{PFV}.} are given by ${\bf p}_{\rm t} \in \left\{ \frac{1}{\sqrt{2}}[1,j]^T, \frac{1}{\sqrt{2}}[1,-j]^T \right\}$ and ${\bf p}_{\rm r} \in \left\{ [1,j]^T, [1,-j]^T \right\}$, respectively.

	\item \textbf{PAA}: This system is equipped with $N$ transmit and $M$ receive PAAs, where each PAA can dynamically adjust its polarization angle within the range of $0$ to $2\pi$.  The polarization vectors are defined as ${\bf p}_{\rm t} = [\cos\alpha, \sin\alpha]^T$ and ${\bf p}_{\rm r} = [\cos\beta, \sin\beta]^T$ \cite{ref_KM15,ref_CH24}, with $\alpha$ and $\beta$ being polarization angles at the transmitter and receiver, respectively.
	
	\item \textbf{CPA}: The polarization of all $N$ transmit and $M$ receive antennas is fixed to left-handed circular polarization, with the polarization vectors given by ${\bf p}_{\rm t} = \frac{1}{\sqrt{2}}[1,j]^T$ for the transmitter and ${\bf p}_{\rm r} = [1,j]^T$ for the receiver.
	
	\item \textbf{LPA}: The polarization of all $N$ transmit and $M$ receive antennas is fixed to vertical polarization, with the polarization vectors given by ${\bf p}_{\rm t} = [1,0]^T$ for the transmitter and ${\bf p}_{\rm r} = [1,0]^T$ for the receiver.
\end{itemize}
For the above benchmark schemes, the polarization vector at the transmitter is normalized to satisfy the transmit power constraint, but normalization at the receiver is not necessary.

Fig.\;\ref{fig_compare_DPA} compares polarforming applied at both the transmitter and receiver with the DPA scheme for different numbers of antennas.  In this figure, for the same number of antennas, the proposed scheme achieves higher rates than the DPA scheme in the low SNR regime but underperforms in the high SNR regime.  This occurs because the DPA scheme benefits from double data streams, which provide a higher multiplexing gain; however, this advantage diminishes when noise power dominates.  On the other hand, the proposed scheme outperforms the DPA scheme under the same number of RF chains, where the DPA scheme has half the number of antennas compared to the proposed scheme.  { Specifically, the proposed scheme achieves SNR gains of $6.8$ dB at the rate of $10$ bps/Hz, $6.1$ dB at the rate of $20$ bps/Hz, and $5.2$ dB at the rate of $30$ bps/Hz over the DPA scheme for two, four, and eight RF chains, respectively.}

Fig.\;\ref{fig_PF} shows the performance of transmit/receive polarforming in the MISO/SIMO system when it is applied at the transmitter/receiver only.  In transmit polarforming (see Figs. \ref{fig_txp_LPA} and \ref{fig_txp_CPA}), an LPA or CPA is deployed at the receiver, while two PS-based PRAs are deployed at the transmitter.  Conversely, in receive polarforming (see Figs. \ref{fig_rxp_LPA} and \ref{fig_rxp_CPA}), an LPA or CPA is deployed at the transmitter, and two PS-based PRAs are deployed at the receiver.  The optimal solution for transmit/receive polarforming is based on \eqref{opt_theta_miso} and \eqref{opt_phi_simo} as discussed in Section \ref{sec3-2}.  From Fig.\;\ref{fig_txp_LPA}, the schemes using antennas with a single element, such as the PAA and LPA schemes, perform better than the proposed and CPA schemes, respectively.  However, this advantage diminishes when a CPA is used at the receiver, as shown in Fig.\;\ref{fig_txp_CPA}.  This occurs because the antennas with dual elements lose their advantage under the transmit power constraint. Additionally, the schemes using a single element achieve better performance when the inverse XPD is set to $\chi = 0.2$, which corresponds to weak channel depolarization.  In contrast, receive polarforming generally achieves higher rates than transmit polarforming, as shown in Figs. \ref{fig_rxp_LPA} and \ref{fig_rxp_CPA}.  Moreover, the antennas with dual elements outperform those with a single element, such that the proposed scheme surpasses the PAA scheme, and the CPA scheme outperforms the LPA scheme. Notably, flexible-polarization antennas consistently perform better than FPAs, i.e., the proposed and SPRA schemes outperform the CPA scheme, and the PAA scheme outperforms the LPA scheme.  The results in these figures demonstrate that the proposed polarforming scheme generally achieves better performance as compared to benchmark schemes, particularly at the receiver, where it outperforms the PAA scheme.

\begin{figure}[!t]
	\centering
	\includegraphics[width=0.9\columnwidth]{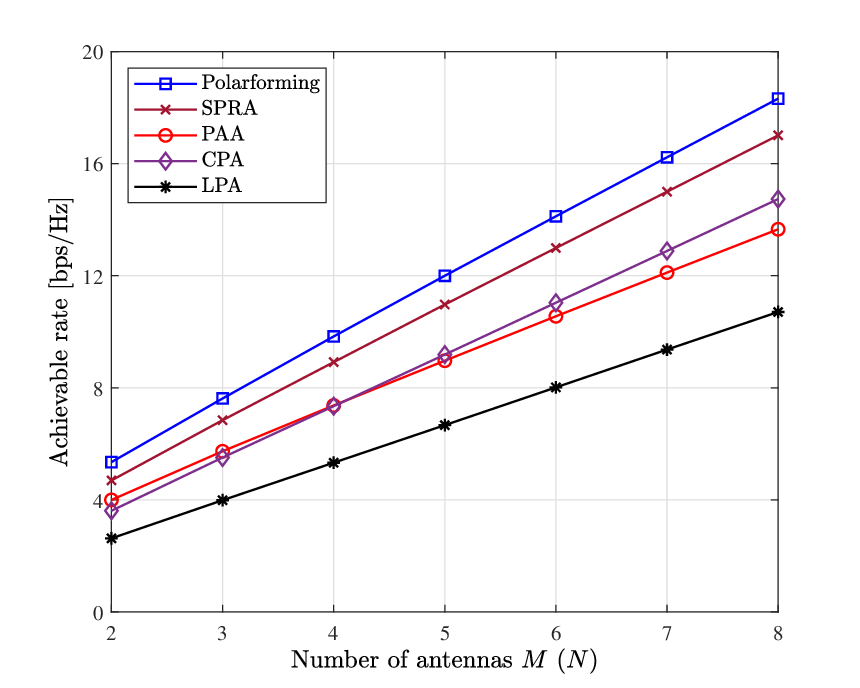}
	\caption{{ Achievable rate of polarforming versus number of antennas.}}
	\label{fig_impact_ant}
\end{figure}

Fig.\;\ref{fig_rate_snr} depicts the achievable rates of polarforming applied at both the transmitter and receiver and benchmark schemes versus SNR for different scenarios.  The simulations for the SISO, MISO, and SIMO scenarios are conducted using Algorithm \ref{alg1}, while Algorithm \ref{alg2} is used for the MIMO scenario.  From these figures, it is evident that the proposed scheme achieves superior achievable rates compared to all benchmark schemes, with the performance gap remaining nearly consistent as SNR increases.  Specifically, at the rate of $4$ bps/Hz, the proposed scheme achieves SNR gains of $1.9$ dB, $2.7$ dB, $5.6$ dB, and $6.3$ dB over the conventional SPRA, PAA, CPA, and LPA schemes, respectively, for $M=N=1$.  Similarly, for $M=N=2$, the corresponding SNR gains are $1.4$ dB, $2.8$ dB, $4.1$ dB, and $4.8$ dB.  Notably, due to channel reciprocity, the performance of the mentioned schemes for the MISO ($M=1,N=2$) and SIMO ($M=2,N=1$) scenarios is identical, with the proposed scheme achieving SNR gains of $1.6$ dB, $2.7$ dB, $4.5$ dB, and $5.3$ dB.  These results demonstrate the significant advantage of polarforming in enhancing communication performance.

Fig.\;\ref{fig_impact_ant} illustrates the achievable rates of polarforming applied at both the transmitter and receiver and benchmark schemes for different numbers of antennas by assuming $M=N$ at $\text{SNR} = 5$ dB.  This figure confirms that the proposed scheme consistently outperforms the schemes using conventional PRAs and FPAs across various numbers of antennas.  Interestingly, the CPA scheme outperforms the PAA scheme as the number of antennas increases, in particular, for $M (N) \ge 5$.  This suggests that CPAs benefit more from enhanced spatial diversity, whereas the performance of PAAs is constrained due to the use of LPAs.

\begin{figure}[!t]
	\centering
	\includegraphics[width=0.9\columnwidth]{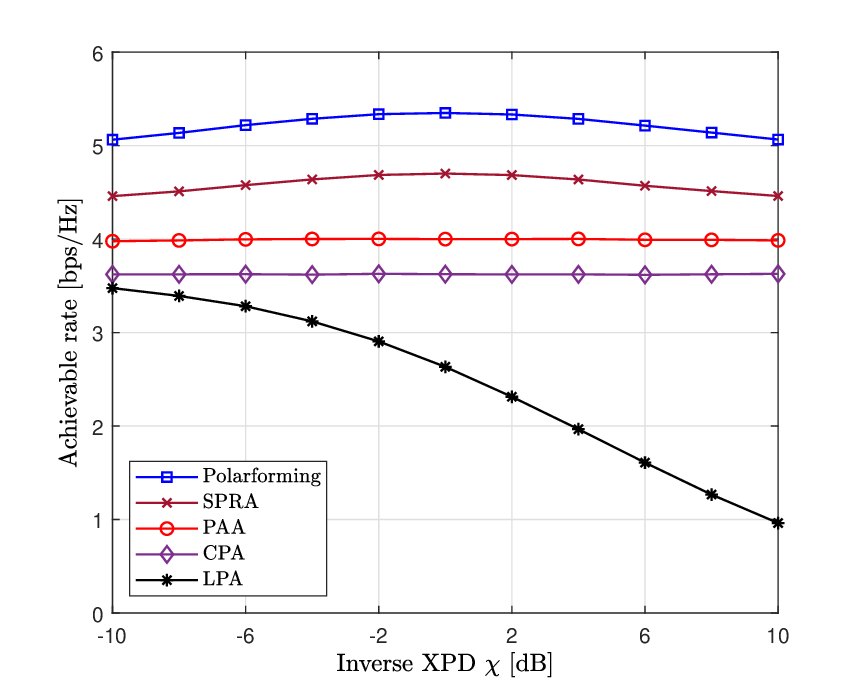}
	\caption{{ Performance of polarforming against channel depolarization.}}
	\label{fig_impact_xpd}
\end{figure}

\begin{figure}[!t]
	\centering
	\subfigure[Antenna XPI]{\label{fig_impact_xpi}
		\centering
		\includegraphics[width=0.9\columnwidth]{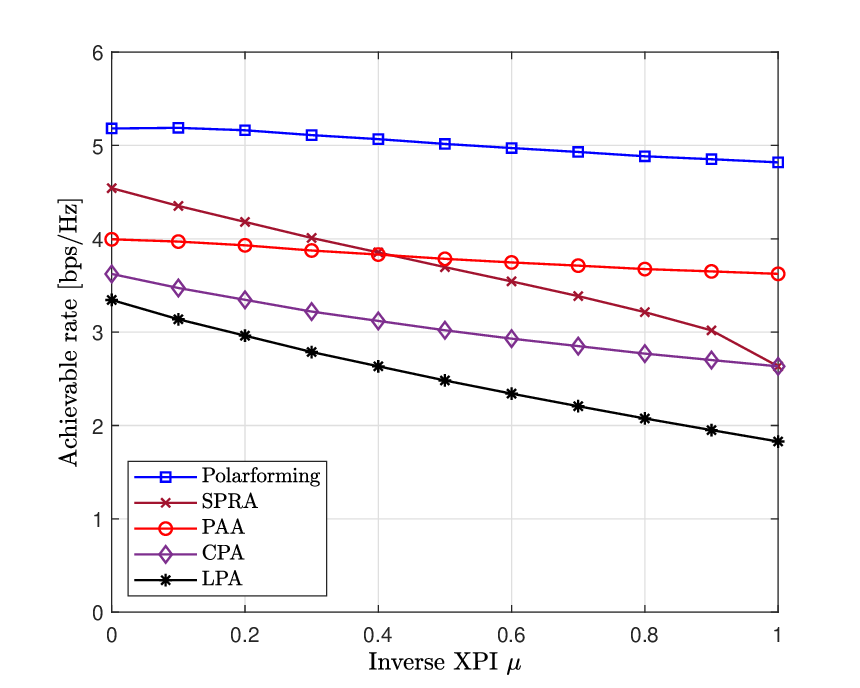}}
	\subfigure[Polarized channel correlation]{\label{fig_impact_corr}
		\centering
		\includegraphics[width=0.9\columnwidth]{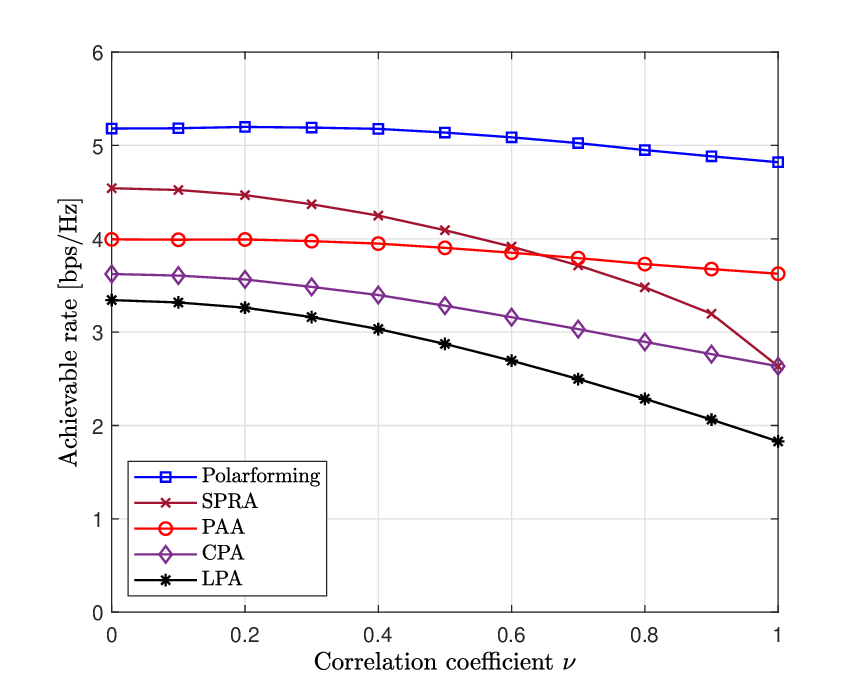}}
	\subfigure[{ Polarized channel estimation error}]{\label{fig_impact_err}
		\centering
		\includegraphics[width=0.9\columnwidth]{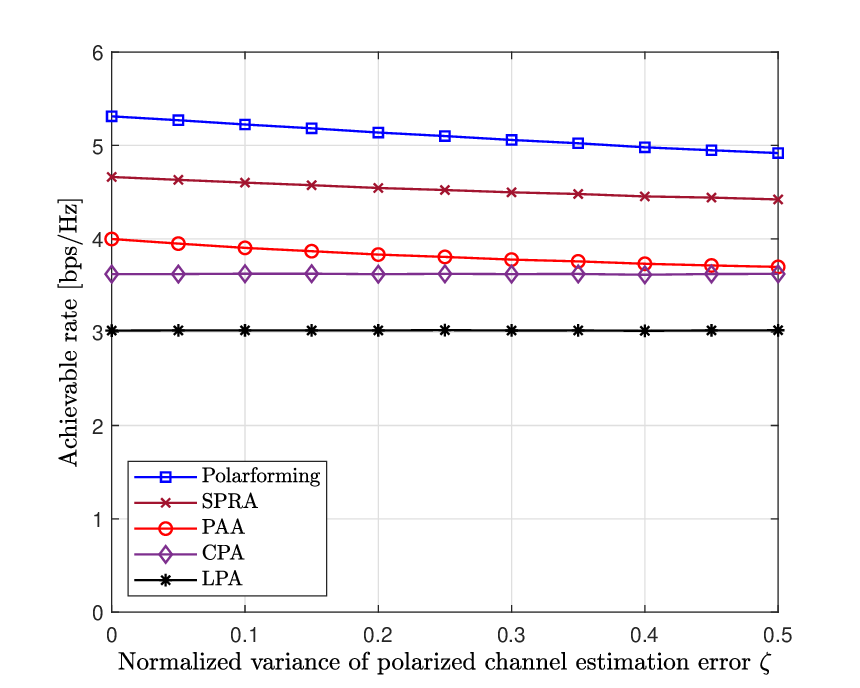}}
	\caption{{ Impacts of antenna XPI, polarized channel correlation, and polarized channel error on the performance of polarforming.}}
	\label{fig_impact}
\end{figure}

Fig.\;\ref{fig_impact_xpd} displays the performance of polarforming applied at both the transmitter and receiver against channel depolarization for $M=N=2$ and $\text{SNR} = 5$ dB.  It is worth noting that the non-decreasing rate versus increasing inverse XPD indicates better robustness against channel depolarization.  The proposed scheme is observed to outperform conventional schemes using SPRAs, PAAs, and FPAs, which suggests that polarforming is more effective for combating channel depolarization across varying inverse XPD values.  Moreover, the maximum achievable rate of the proposed scheme is attained when the inverse XPD equals one.  This occurs because the average amplitudes in the cross-polar and co-polar channels are equal.  In other words, the polarization angle determined by the equal amplitudes happens to align with the fixed polarization angle of PS-based PRAs, as the V-element and H-element of each PRA are designed with identical fixed amplitudes.  Since the considered SPRA scheme can be regarded as a special case of polarforming with discrete phase shifts, it exhibits a similar trend to polarforming.  Additionally, the CPA scheme demonstrates a flat rate curve with respect to inverse XPD, while the rate of the LPA scheme decreases as inverse XPD increases.  This result is due to the different characteristics of CPAs and LPAs, i.e., the orthogonal antenna elements of CPAs enable channel power acquisition in two orientations, whereas LPAs are limited to a single orientation, as demonstrated in \cite{ref_DGM17}.

{ Fig.\;\ref{fig_impact} further investigates the impacts of antenna XPI, polarized channel correlation, and polarized channel estimation error on the performance of polarforming with $M=N=2$ and $\text{SNR} = 5$ dB.}  In Figs. \ref{fig_impact_xpi} and \ref{fig_impact_corr}, the simulation setup is based on the channel models given in Section \ref{sec3-4}, with the parameters $\mu_{\rm t} = \mu_{\rm r} \triangleq \mu$ and $\nu_{\rm t} = \nu_{\rm r} \triangleq \nu$.  Since higher antenna XPI is equivalent to increasing the polarized channel correlation, the results in these figures exhibit similar trends.  It can be seen that all schemes experience performance degradation as the inverse XPI or correlation coefficient increases.  This degradation occurs because channel decorrelation reduces the system's ability to effectively separate signals, thereby deteriorating the performance of a polarized system.   However, the proposed scheme exhibits a slower degradation trend when $\mu$ or $\nu$ is small and outperforms the benchmark schemes.   Interestingly, the SPRA and CPA schemes converge at the same rate.  This is owing to $\mu = 1$ or $\nu = 1$, which indicates the polarized channels are fully correlated, and switching between two polarization states loses its effectiveness in this case.  { In Fig.\;\ref{fig_impact_err}, the polarized channel estimation error is quantified by the element-wise difference between the channel matrix ${\bf P}$ and its estimate $\hat{\bf P}$, expressed as $\frac{[{\bf P}]_{kl} - [\hat{\bf P}]_{kl}}{|[{\bf P}]_{kl}|} \sim {\cal CN}(0, \zeta)$ for $1\le k\le 2M$ and $1\le l \le 2N$, where $\zeta$ denotes the normalized error variance.  The results indicate that the proposed scheme degrades gradually as $\zeta$ increases, yet retains a consistent performance advantage over all benchmark methods throughout the error range.  Even under large estimation errors, the proposed scheme exhibits only moderate degradation, which underscores its robustness to imperfect polarized channel estimation.  In contrast, the PAA scheme suffers from faster performance decay due to its stronger dependence on precise channel knowledge for effective transmission.  Meanwhile, the LPA and CPA schemes show relatively stable but significantly lower performance with varying $\zeta$, confirming their limited ability to exploit channel knowledge for polarization adaptation.}

\begin{figure}[!t]
	\centering
	\includegraphics[width=0.9\columnwidth]{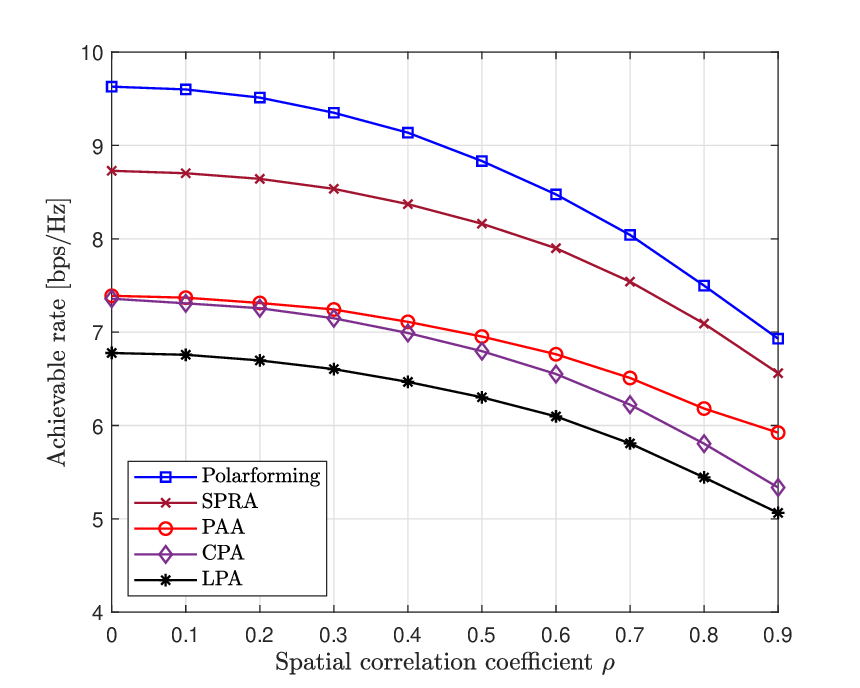}
	\caption{Impact of ULA-based spatial correlation on the performance of polarforming.}
	\label{fig_impact_ula}
\end{figure}

Beyond the i.i.d. polarized channels, we further examine the robustness of polarforming under spatially correlated channels with $M=N=4$ and $\text{SNR} = 5$ dB.  Following the widely adopted uniform linear array (ULA) model, the spatial correlation across antennas is characterized by the exponential correlation matrices ${\bf R}_{\rm t}$ and ${\bf R}_{\rm r}$ with entries $[{\bf R}]_{ik} = \rho^{|i-k|}$ \cite{ref_Loyka01}, where $\rho \in [0,1)$ denotes the spatial correlation coefficient.  The correlated polarized channel matrix is generated as ${\bf P} = ({\bf R}_{\rm r} \otimes {\bf I}_2)^{1/2}\, {\bf P}^{\rm i.i.d.}\, ({\bf R}_{\rm t} \otimes {\bf I}_2)^{1/2}$, where ${\bf P}^{\rm i.i.d.}$ denotes the i.i.d. polarized channel matrix generated as above, $\otimes$ denotes the Kronecker product, and the polarization structure of each antenna pair is preserved.  Fig.\;\ref{fig_impact_ula} illustrates the achievable rate versus the spatial correlation coefficient $\rho$.  As expected, the rates of all schemes decrease as $\rho$ increases owing to the reduced spatial multiplexing gain.  Nevertheless, the proposed polarforming scheme maintains a consistent performance advantage over the SPRA, PAA, CPA, and LPA benchmarks across the entire range of spatial correlation.  Specifically, it outperforms the PAA scheme by about $2.2$ bps/Hz and $1.0$ bps/Hz at $\rho = 0$ and $\rho = 0.9$, respectively.  This confirms that the polarization-domain gains of polarforming are present even under highly correlated ULA channels.

\section{Conclusion} \label{sec5}
In this paper, we introduced and investigated the new concept of polarforming that enables antennas to configure a desired polarization state and thereby adjust the polarization of transmitted/received EM waves.  We proposed a novel PS-based PRA and studied a single-user MIMO system aided by this type of PRA.  With the PS-based PRA, we provided a detailed illustration of polarforming and explained its differences from conventional beamforming.  Furthermore, for the proposed system, we presented an alternating optimization approach to maximize the channel capacity for the MISO and SIMO scenarios.  We also derived an upper bound for the MIMO channel capacity and developed an iterative algorithm to optimize the phase shifts of PRAs by maximizing this upper bound.  Furthermore, the simulation results validated and demonstrated the effectiveness of the PS-based PRA system.  It was shown that PS-based PRAs outperform conventional PRAs and FPAs in combating channel depolarization and adapting to channel variations.  Besides, PS-based PRAs perform better than DPAs when using the same number of RF chains or in the low SNR regime with the same number of antennas.  This implies that PS-based PRAs can serve as an alternative approach to exploiting polarization diversity by reducing the number of required RF chains compared to DPA systems.  In future work, we will investigate polarforming under measured polarized channels and wideband scenarios, as well as its joint design with beamforming in different PRA-based wireless systems.


\begin{thebibliography}{1}
	\bibliographystyle{IEEEtran}
	\bibitem{ref_AMD01}
	M. R. Andrews, P. P. Mitra, and R. deCarvalho, ``Tripling the capacity of wireless communications using electromagnetic polarization,'' \textit{Nature}, vol. 409, pp. 316–318, Jan. 2001.
	
	\bibitem{ref_AOK24}
	M. Aldababsa, S. {\"O}zyurt, G. K. Kurt, and O. Kucur, ``A survey on orthogonal time frequency space modulation,'' \textit{IEEE Open J. Commun. Soc.}, vol. 5, pp. 4483-4518, 2024.
	
	\bibitem{ref_RXK19}
	T. S. Rappaport, Y. Xing, O. Kanhere, S. Ju, A. Madanayake, S. Mandal, A. Alkhateeb, and G. C. Trichopoulos, ``Wireless communications and applications above 100 Ghz: Opportunities and challenges for 6G and beyond,'' \textit{IEEE Access}, vol. 7, pp. 78729–78757, 2019.
	
	\bibitem{ref_LLS14}
	L. Lu, G. Y. Li, A. L. Swindlehurst, A. Ashikhmin, and R. Zhang, ``An overview of massive MIMO: Benefits and challenges,'' \textit{IEEE J. Sel. Topics Signal Process.}, vol. 8, no. 5, pp. 742–758, Oct. 2014.

	\bibitem{ref_MO09}
	F. Mani and C. Oestges, ``A ray based indoor propagation model including depolarizing penetration,'' in \textit{Proc. 3rd Eur. Conf. Antennas and Propag.}, Mar. 2009, pp. 3835–3838.
	
	\bibitem{ref_KS11}
	S.-C. Kwon and G. L. Stuber, ``Geometrical theory of channel depolarization,'' \textit{IEEE Trans. Veh. Technol.}, vol. 60, no. 8, pp. 3542–3556, Oct.	2011.
	
	\bibitem{ref_Coldrey08}
	M. Coldrey, ``Modeling and capacity of polarized MIMO channels,'' in \textit{Proc. VTC Spring}, May 2008, pp. 440–444.
	
	\bibitem{ref_HCS16}
	Y. He, X. Cheng, and G. L. St{\"u}ber, ``On polarization channel modeling,'' \textit{IEEE Wireless Commun.}, vol. 23, no. 1, pp. 80-86, Feb. 2016.
	
	\bibitem{ref_NBE02}
	R. U. Nabar, H. Bolcskei, V. Erceg, D. Gesbert, and A. J. Paulraj, ``Performance of multiantenna signaling techniques in the presence of polarization diversity,'' \textit{IEEE Trans. Signal Process.}, vol. 50, no. 10,	pp. 2553–2562, Oct. 2002.
	
	\bibitem{ref_ESB04}
	V. Erceg, P. Soma, D. Baum, and S. Catreux, ``Multiple-input multiple-output fixed wireless radio channel measurements and modeling using dual-polarized antennas at 2.5 GHz,'' \textit{IEEE Trans. Wireless Commun.}, vol. 3, no. 6, pp. 2288–2298, Nov. 2004.
	
	\bibitem{ref_ESC06}
	V. Eiceg, H. Sampath, and S. Catreux-Erceg, ``Dual-polarization versus single-polarization MIMO channel measurement results and modeling,'' \textit{IEEE Trans. Wireless Commun.}, vol. 5, no. 1, pp. 28-33, Jan. 2006.
	
	\bibitem{ref_ABH07}
	H. Asplund, J.-E. Berg, F. Harrysson, J. Medbo, and M. Riback, ``Propagation characteristics of polarized radio waves in cellular communications,'' in \textit{Proc. IEEE VTC}, Baltimore, MD, USA, 2007,	pp. 839–843.
	
	\bibitem{ref_VGM09}
	J. F. Valenzuela-Valdes, M. A. Garcia-Fernandez, A. M. Martinez-Gonzalez, and D. A. Sanchez-Hernandez, ``Evaluation of true polarization diversity for MIMO systems,'' \textit{IEEE Trans. Antennas Propag.}, vol. 57, no. 9, pp. 2746-2755, Sep. 2009.

	\bibitem{ref_YSL23}
	Y. Yao, F. Shu, Z. Li, X. Cheng, and L. Wu, ``Secure transmission scheme based on joint radar and communication in mobile vehicular networks,''	\textit{IEEE Trans. Intell. Transp. Syst.}, vol. 24, no. 9, pp. 10027–10037, Sep. 2023.

	\bibitem{ref_MWX24}
	K. Meng, Q. Wu, J. Xu, W. Chen, Z. Feng, R. Schober, and A.	L. Swindlehurst, ``UAV-enabled integrated sensing and communication:	Opportunities and challenges,'' \textit{IEEE Wireless Commun.}, vol. 31, no. 2, pp. 97–104, Jul. 2024.
	
	\bibitem{ref_OB23}
	\"O. \"Ozdogan and E. Bj\"ornson, ``Massive MIMO with dual-polarized antennas,'' \textit{IEEE Trans. Wireless Commun.}, vol. 22, no. 2, pp. 1448-1463, Feb. 2023.

	\bibitem{ref_OCG08}
	C. Oestges, B. Clerckx, M. Guillaud, and M. Debbah, ``Dual-polarized wireless communications: From propagation models to system performance evaluation,'' \textit{IEEE Trans. Wireless Commun.}, vol. 7, no. 10, pp. 4019-4031, Oct. 2008.
	
	\bibitem{ref_Ertug08}
	O. Ertug, ``Asymptotic ergodic capacity of multidimensional vector-sensor array MIMO channels,'' \textit{IEEE Trans. Wireless Commun.}, vol. 7, no. 9, pp. 3297-3300, Sep. 2008.
	
	\bibitem{ref_IKB24}
	A. Irshad, A. Kosasih, E. Bj\"ornson, and L. Sanguinetti, ``Optimal dual-polarized planar arrays for massive capacity over point-to-point MIMO channels,'' \textit{IEEE Trans. Wireless Commun.}, vol. 23, no. 12, pp. 19350-19364, Dec. 2024.
	
	\bibitem{ref_WPF19}
	X. Wu, T. G. Pratt, and T. E. Fuja, ``Polarization shift keying for wireless communication,'' \textit{IEEE Trans. Wireless Commun.}, vol. 18, no. 10, pp. 4927-4941, Oct. 2019.
	
	\bibitem{ref_WPF20}
	X. Wu, T. G. Pratt, and T. E. Fuja, ``Hybrid constellations for dual-polarized wireless communications,'' \textit{IEEE Trans. Wireless Commun.}, vol. 19, no. 8, pp. 5321-5332, Aug. 2020.
	
	\bibitem{ref_HXK20}
	I. A. Hemadeh, P. Xiao, Y. Kabiri, L. Xiao, V. Fusco, and R. Tafazolli, ``Polarization modulation design for reduced RF chain wireless,'' \textit{IEEE Trans. Commun.}, vol. 68, no. 6, pp. 3890-3907, Jun. 2020.

	\bibitem{ref_XHL14}
	L. D. Xu, W. He, and S. Li, ``Internet of Things in industries: A survey,'' \textit{IEEE Trans. Ind. Informat.}, vol. 10, no. 4, pp. 2233–2243, Nov. 2014.
		
	\bibitem{ref_SCL15}
	J. Song, J. Choi, S. G. Larew, D. J. Love, T. A. Thomas, and A. A. Ghosh, ``Adaptive millimeter wave beam alignment for dual-polarized MIMO systems,'' \textit{IEEE Trans. Wireless Commun.}, vol. 14, no. 11, pp. 6283-6296, Nov. 2015.
	
	\bibitem{ref_ZCH17}
	D. Zhu, J. Choi and R. W. Heath, ``Two-dimensional AoD and AoA acquisition for wideband millimeter-wave systems with dual-polarized MIMO,'' \textit{IEEE Trans. Wireless Commun.}, vol. 16, no. 12, pp. 7890-7905, Dec. 2017.
	
	\bibitem{ref_SJK04}
	Y. J. Sung, T. U. Jang, and Y.-S. Kim, ``A reconfigurable microstrip antenna for switchable polarization,'' \textit{IEEE Microw. Wireless Compon.	Lett.}, vol. 14, no. 11, pp. 534–536, Nov. 2004.
	
	\bibitem{ref_GSZ06}
	S. Gao, A. Sambell, and S. S. Zhong, ``Polarization-agile antennas,'' \textit{IEEE Antennas Propag. Mag.}, vol. 48, no. 3, pp. 28-37, Jun. 2006.
	
	\bibitem{ref_KRR15}
	J. M. Kovitz, H. Rajagopalan, and Y. Rahmat-Samii, ``Design and implementation of broadband MEMS RHCP/LHCP reconfigurable arrays using rotated E-shaped patch elements,'' \textit{IEEE Trans. Antennas Propag.}, vol. 63, no. 6, pp. 2497–2507, Jun. 2015.
	
	\bibitem{ref_ZCL15}
	H. L. Zhu, S. W. Cheung, X. H. Liu, and T. I. Yuk, ``Design of polarization reconfigurable antenna using metasurface,'' \textit{IEEE Trans. Antennas Propag.}, vol. 62, no. 6, pp. 2891-2898, Jun. 2014.
	
	\bibitem{ref_RZW20}
	J. Ren, Z. Zhou, Z. H. Wei, H. M. Ren, Z. Chen, Y. Liu, and Y. Z. Yin, ``Radiation pattern and polarization reconfigurable antenna using dielectric liquid,'' \textit{IEEE Trans. Antennas Propag.}, vol. 68, no. 12, pp. 8174–8179, Dec. 2020.

	\bibitem{ref_KM15}
	S.-C. Kwon and A. F. Molisch, ``Capacity maximization with polarization-agile antennas in the MIMO communication system,'' in \textit{Proc. IEEE Global Commun. Conf. (GLOBECOM)}, San Diego, CA, USA, Dec. 2015, pp. 1–6.
	
	\bibitem{ref_OKY18}
	S. P. Oh and S.-C. S. Kwon, ``Capacity of polarized-MIMO (P-MIMO) system in different wireless channels,'' in \textit{Proc. IEEE Green Energy Smart Syst. Conf.	(IGESSC)}, Long Beach, CA, USA, 2018, pp. 1-6.
	
	\bibitem{ref_OK20}
	P. Oh and S. Kwon, ``Multi-polarization superposition beamforming with XPD-aware transmit power allocation,'' in \textit{Proc. IEEE 92nd Veh. Technol. Conf. (VTC-Fall)}, Victoria, BC, Canada, Nov. 2020, pp. 1–6.
	
	\bibitem{ref_OKM21}
	P. S. Oh, S. S. Kwon, and A. F. Molisch, ``Antenna selection in polarization reconfigurable MIMO (PR-MIMO) communication systems,'' arXiv preprint \textit{arXiv:2112.00931}, 2021.

	\bibitem{ref_OHK24}
	S. Oh, H. Han, J. Kim, and S. Kwon, ``Double-side polarization and beamforming alignment in polarization reconfigurable MISO system with deep neural networks,'' arXiv preprint \textit{arXiv:2409.20065}, 2024.
	
	\bibitem{ref_CH21}
	M. R. Castellanos and R. W. Heath, ``MIMO communication with polarization reconfigurable antennas,'' in \textit{Proc. 55th Asilomar Conf. Signals, Syst., Comput.}, Pacific Grove, CA, USA, Oct. 2021, pp. 1–5.
	
	\bibitem{ref_CH24}
	M. R. Castellanos and R. W. Heath, ``Linear polarization optimization for wideband MIMO systems with reconfigurable arrays,'' \textit{IEEE Trans. Wireless Commun.}, vol. 23, no. 3, pp. 2282-2295, Mar. 2024.
		
	\bibitem{ref_DKY19}
	S. Doan, S. Kwon, and H.-G. Yeh, ``Achievable capacity of multi-polarization MIMO (MP-MIMO) toward 6G wireless communications,'' in \textit{Proc. IEEE Green Energy Smart Syst. Conf.	(IGESSC)}, Long Beach, CA, USA, 2019, pp. 1-6.
	
	\bibitem{ref_DKY21}
	S. H. Doan, S.-C. Kwon, and H.-G. Yeh, ``Achievable capacity of multipolarization MIMO with the practical polarization-agile antennas,'' \textit{IEEE Syst. J.}, vol. 15, no. 2, pp. 3081–3092, Jun. 2021.
	
	\bibitem{ref_DGM17}
	F. A. Dicandia, S. Genovesi, and A. Monorchio, ``Analysis of the performance enhancement of MIMO systems employing circular polarization,'' \textit{IEEE Trans. Antennas Propag.}, vol. 65, no. 9, pp. 4824–4835, Sep. 2017.
	
	\bibitem{ref_ZZ20}
	S. Zhang and R. Zhang, ``Capacity characterization for intelligent reflecting surface aided MIMO communication,'' \textit{IEEE J. Sel. Areas Commun.}, vol. 38, no. 8, pp. 1823–1838, Aug. 2020.
	
	\bibitem{ref_MZZ23}
	W. Ma, L. Zhu, and R. Zhang, ``MIMO capacity characterization for movable antenna systems,'' \textit{IEEE Trans. Wireless Commun.}, vol. 23, no. 4, pp. 3392-3407, Apr. 2024.
	
	\bibitem{ref_TV05}
	D. Tse and P. Viswanath, \textit{Fundamentals of Wireless Communication}. Cambridge, U.K.: Cambridge Univ. Press, 2005.
	
	\bibitem{ref_Balanis16}
	C. A. Balanis, \textit{Antenna Theory: Analysis and Design}, 4th ed. Hoboken, NJ, USA: Wiley, 2016.
	
	\bibitem{ref_CP12}
	J. Chen and T. G. Pratt, ``A three-dimensional geometry-based statistical model of 2 $\times$ 2 dual-polarized MIMO mobile-to-mobile wideband channels,'' \textit{Model. Simul. Eng.}, vol. 2012, no. 756508, pp. 1–16, Nov. 2012.

	\bibitem{ref_ZSE22}
  H. Zhang, N. Shlezinger, F. Guidi, D. Dardari, M. F. Imani, and Y. C. Eldar, ``Beam focusing for near-field multiuser MIMO communications,'' \textit{IEEE Trans. Wireless Commun.}, vol. 21, no. 9, pp. 7476–7490, Sep. 2022.

	\bibitem{ref_LMD23}
	Y. Liu, Z. Wang, J. Xu, C. Ouyang, X. Mu, and R. Schober, ``Near-field communications: A tutorial review,'' \textit{IEEE Open J. Commun. Soc.}, vol. 4, pp. 1999–2049, 2023.

	\bibitem{ref_Loyka01}
	S. Loyka, ``Channel capacity of MIMO architecture using the exponential correlation matrix,'' \textit{IEEE Commun. Lett.}, vol. 5, no. 9, pp. 369--371, Sep. 2001.

\end{thebibliography}
\end{document}